\documentclass[useAMS]{mn2e}

\usepackage{graphicx}
\usepackage{amssymb}
\usepackage{amsmath}
\title[Radio spectra of luminous IRAS galaxies]{Starburst evolution: free-free absorption in the radio spectra of luminous IRAS galaxies}
\author[M. S. Clemens, A. Scaife, O. Vega, A. Bressan]{M. S. Clemens,$^{1}$\thanks{E-mail:
marcel.clemens@oapd.inaf.it (MSC); alessandro.bressan@oapd.inaf.it (AB)} A. Scaife$^{2,5}$, O. Vega$^{3}$, A.
Bressan$^{1,3,4}$\footnotemark[1]\\
$^{1}$INAF-Osservatorio Astronomico di Padova, Vicolo dell'Osservatorio, 5, 35122 Padova, Italy\\
$^{2}$Astrophysics Group, Cavendish Laboratory, J. J. Thomson Avenue, Cambridge CB3 0HE, UK \\
$^{3}$INAOE, Luis Enrique Erro 1, 72840 Tonantzintla, Puebla, Mexico\\
$^{4}$SISSA-ISAS, International School for Advanced Studies, ia Beirut 4, 34014 Trieste, Italy\\
$^{5}$Dublin Institute for Advanced Studies, 31 Fitzwilliam Place, Dublin 2,Ireland }

\begin{document}

\date{Accepted ????. Received ????; in original form ????}

\pagerange{\pageref{firstpage}--\pageref{lastpage}} \pubyear{2009}

\maketitle

\label{firstpage}

\begin{abstract}
We describe radio observations at 244 and $610\;\rm MHz$ of a sample of 20 luminous and ultra-luminous IRAS galaxies. These are a sub-set of a sample of 31 objects that have well-measured radio spectra up to at least $23\;\rm GHz$. The radio spectra of these objects below $1.4\;\rm GHz$ show a great variety of forms and are rarely a simple power-law extrapolation of the synchrotron spectra at higher frequencies. Most objects of this class have spectral turn-overs or bends in their radio spectra. We interpret these spectra in terms of free-free absorption in the starburst environment. 

Several objects show radio spectra with two components having free-free turn-overs at different frequencies (including Arp~220 and Arp~299), indicating that synchrotron emission originates from regions with very different emission measures. In these sources, using a simple model for the supernova rate, we estimate the time for which synchrotron emission is subject to strong free-free absorption by ionized gas, and compare this to expected HII region lifetimes. We find that the ionized gas lifetimes are an order of magnitude larger than plausible lifetimes for individual HII regions. We discuss the implications of this result and argue that those sources which have a significant radio component with strong free-free absorption are those in which the star formation rate is still increasing with time. 

We note that if ionization losses modify the intrinsic synchrotron spectrum so that it steepens toward higher frequencies, the often observed deficit in fluxes higher than $\sim 10\;\rm GHz$ would be much reduced.

%if the star formation is exponentially decreasing with time. Neither metallicity nor IMF variations stronly influence this result. Rising star formation rates, however, greatly reduce the derived lifetimes because the number of massive, short-lived stars, which explode as supernovae while still embedded in HII regions is increased with respect to those of lower mass that out-live HII regions. This is analogous to the age-dependent extinction already noted at other wavelengths. For this effect to be efficient requires starburst rise times to be much shorter than the, $\simeq 50\;\rm Myr$, lifetime of the least massive stars that produce type II supernovae. We argue that those sources which have a significant radio component with strong free-free absorption are those in which the star formation rate is still increasing with time. 

%We note that if ionization losses modify the intrinsic synchrotron spectrum so that it steepens toward higher frequencies, the often observed deficit in fluxes higher than $\sim 10\;\rm GHz$ would be much reduced.

%The radio emission from the AGN in UGC~8058 (Mrk~231) is separated from that of the star formation because the two components have different turn-over frequencies. At the epoch of the data above 1.4~GHz, the AGN contributed 66\% of the radio emission.   

\end{abstract}

\begin{keywords}
Galaxies: active, Infrared: galaxies, Radio continuum: galaxies
\end{keywords}

\section{Introduction}

\begin{table}
\caption{Total 244 and $610\;\rm MHz$ fluxes. Distances are as reported in Condon et al. (1991).}
\centering
\begin{tabular}{c c c c}
\hline\hline
Source          &  D   & $f_{244}$         & $f_{610}$ \\
                & (Mpc)& (Jy)              & (Jy)  \\
\hline
NGC~34          & 77.1 & ---               & --- \\
IC~1623         & 71.3 & ---               & --- \\
CGCG436-30      & 123.6& ---               & --- \\
IRAS~01364-1042 & 187.2& ---               & --- \\
IIIZw~35        & 108.7& ---               & --- \\
UGC~2369        & 123.5& $0.040 \pm 0.020$ & $0.066 \pm 0.004$ \\
IRAS~03359+1523 & 140.1& $<0.060$          & $0.038 \pm 0.005$ \\
NGC~1614        & 60.9 & $0.250 \pm 0.040$ & $0.202 \pm 0.005$ \\
NGC~2623        & 76.1 & $0.161 \pm 0.030$ & --- \\
IRAS~08572+3915 & 235.8& $<0.050$          & $0.011 \pm 0.006$ \\
UGC~4881        & 162.7& $0.060 \pm 0.010$ & $0.056 \pm 0.007$ \\
UGC~5101        & 164.1& $0.310 \pm 0.030$ & $0.212 \pm 0.004$ \\
IRAS~10173+0828 & 193.7& $<0.030$          & $0.010 \pm 0.002$ \\
IRAS~10565+2448 & 170.0& $0.080 \pm 0.020$ & $0.102 \pm 0.007$ \\
Arp~148         & 142.2& $0.080 \pm 0.020$ & $0.058 \pm 0.005$ \\
UGC~6436        & 139.4& $0.020 \pm 0.010$ & $0.032 \pm 0.004$ \\
Arp~299         & 48.0 & $2.350 \pm 0.090$ & $1.300 \pm 0.030$ \\
IRAS~12112+0305 & 292.1& $0.035 \pm 0.020$ & $0.032 \pm 0.006$ \\
UGC8058         & 172.6& $0.590 \pm 0.020$ & $0.403 \pm 0.020$ \\
UGC~8387        & 96.8 & $0.240 \pm 0.005$ & $0.164 \pm 0.008$ \\
NGC5256         & 115.9& $0.410 \pm 0.040$ & $0.225 \pm 0.006$ \\
UGC~8696        & 157.3& $0.394 \pm 0.020$ & $0.222 \pm 0.008$ \\
IRAS~14348-1447 & 326.7& $<0.060$          & $0.030 \pm 0.006$ \\
IZw~107         & 166.0& $0.180 \pm 0.020$ & $0.090 \pm 0.005$ \\
IRAS~15250+3609 & 218.7& ---               & ---\\
NGC~6286        & 80.3 & $0.500 \pm 0.030$ & $0.309 \pm 0.010$ \\
NGC~7469        & 66.4 & ---               & ---\\
IC~5298         & 110.6& ---               & ---\\
Mrk~331         & 72.3 & ---               & ---\\
\hline
\end{tabular}
\label{tab:fluxes}
\end{table}

The radio emission from star forming galaxies is the result of two emission processes, non-thermal synchrotron, and thermal Bremsstrahlung, or free-free emission. Synchrotron is by far the dominant component in the 1-10~GHz range and is characterized by a power-law emission spectrum, $f_{\nu} \propto \nu^{-\alpha}$ where $\alpha \simeq 0.8$. Free-free emission has an almost flat spectrum and the emission falls with frequency only as $\nu^{-0.1}$. Therefore, towards higher frequencies free-free emission becomes increasingly important and radio spectra are expected to show a flattening beyond $\sim 10\;\rm GHz$. 

In LIRGs and ULIRGs, although such a flattening is sometimes seen, the majority of radio spectra, even when measured to 23~GHz, show no such flattening. In fact, there is a tendency for the spectra to \emph{steepen} (Clemens et al. 2008). Despite this, Clemens et al. (2008) showed that the spectral index between 8.4 and 23~GHz is flatter for sources with high values of the far-infrared-radio flux ratio, $q$. As these sources, with an excess of infrared relative to radio emission, are expected to be younger, they would be expected to have proportionately more free-free gas, and thus shallower spectral indices at high frequencies.   

As we will see later, the radio spectra of LIRGs and ULIRGs also hold surprises towards low frequencies, with spectra showing a variety of bends and turn-overs. From a theoretical standpoint there is more than one way in which the radio spectrum may flatten towards low frequencies. There are those processes that absorb radio photons, free-free absorption and synchrotron self-absorption and those that cause energy losses of the relativistic electron population directly. Of these, ionization losses have been considered by Thompson et al. (2006) as a cause for the flattening of starburst radio spectra towards low frequencies. However, as illustrated by these authors, the flattening is rather gradual. Our data show evidence of rapid changes in spectral index and also inversions of the spectra, much more consistent with an absorption process with a strong energy dependence. 

Both free-free absorption and synchrotron self-absorption could rapidly absorb the power-law synchrotron spectrum toward low frequencies. However, synchrotron self-absorption requires extremely high brightness temperatures in order to be significant, and these are just not attained (with the possible exception of UGC~8058) in our sources (Condon et al. 1991). We therefore consider free-free absorption to be the defining mechanism for the spectral shapes of this sample, especially at low radio frequencies.        

Here we describe new observations of a sample of LIRGs and ULIRGs at 244 and 610~MHz made with the GMRT, which we combine with archival data and our previous 23~GHz data to produce well-sampled radio spectra across this wide frequency range. We consider the structure of the starbursts in these objects that is implied when their radio spectra are described in terms of the free-free absorption of a synchrotron power-law.

\section{Sample selection}

Our sample is based on that of Condon et al. (1991) who made 8.4~GHz observations of the 40 ULIRGs brighter than $5.25\;\rm Jy$ at $60\;\rm \mu m$. In Clemens et al. (2008) we described new observations at 23~GHz and the collection of archival data for a sub-sample of 31 of the 40 described by Condon et al. (1991). Some objects were excluded for observational reasons, as described in Clemens et al. (2008). Only 20 of these 31 objects were observed at the GMRT at 244 and 610~GHz simply due to observing constraints.

\section{Observations}

Observations at 244 and $610\;\rm MHz$ were made between 2008 January 13 and 16 at the Giant Metre-wave Radio Telescope, (GMRT). The two frequencies were observed simultaneously using the `dual' mode, with one polarization for each frequency. The total bandwidth of $8\;(16)\;\rm MHz$ was split into 64 (128) channels at $244\;(610)\;\rm MHz$. Total integration times for the sources were between 30 and 120 minutes with scans on-source being interleaved between scans of phase calibrators approximately every 30 minutes. Both 3C~147 and 3C~286 were used as flux calibrators.

Data reduction was carried out using the NRAO Astronomical Image Processing System {\sc aips}. 
After initial, channel-based, editing to remove data affected by interference, bandpass calibration was applied using one of the flux calibrators. The data were calibrated in the standard way by forming a ``channel 0'' data set from the average of 10 channels near the centre of the band. The complex gains from this process were then copied to the line data before reducing the number of channels by a factor of 10 for the purpose of map making. 

Maps were made by Fourier transformation of the data in each channel to produce 'dirty' images and beams ({\sc aips} task ``imagr''). The images and beams were then combined to produce a single continuum image and beam for each source, thus avoiding bandwidth smearing. The resulting maps were deconvolved using the {\sc clean} algorithm and the combined beam. Where the source was sufficiently bright, self-calibration was further applied (varying only the antenna phases) to improve the signal-to-noise. The final maps typically had resolutions of 13$^{\prime\prime}$ and 5$^{\prime\prime}$ and rms noise levels of $5\;\rm mJy$ and $0.5\;\rm mJy$ at 244 and $610\;\rm MHz$ respectively.     

Flux densities were measured by direct integration on the maps via the {\sc aips} task {\sc tvstat}. 
These integrated fluxes (reported in Table~\ref{tab:fluxes}) have errors which include both the rms noise 
and calibration errors.

\begin{figure*}
%\vspace{-3cm}
\centerline{
\includegraphics[scale = 0.3]{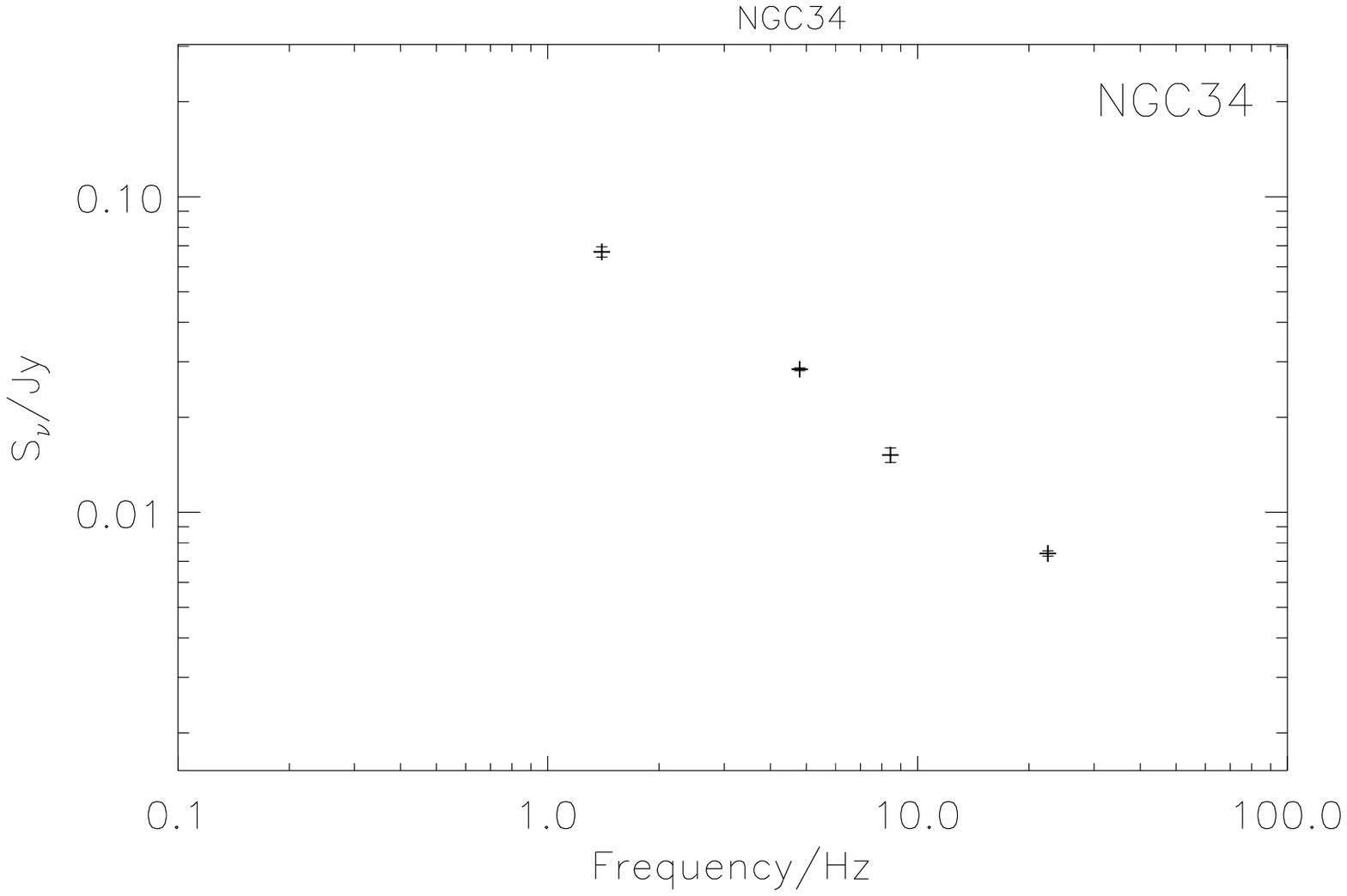}
\includegraphics[scale = 0.3]{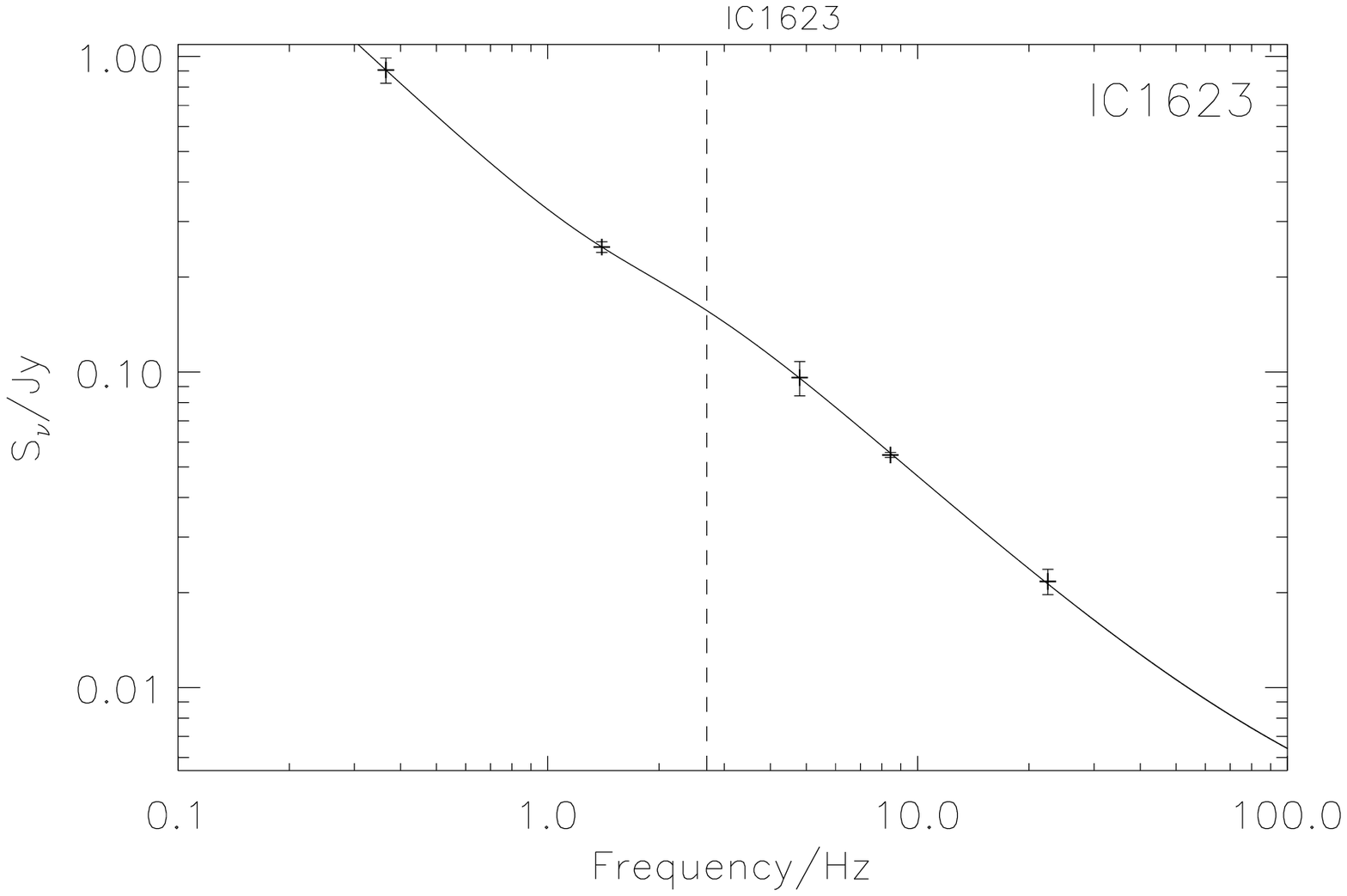}
\includegraphics[scale = 0.3]{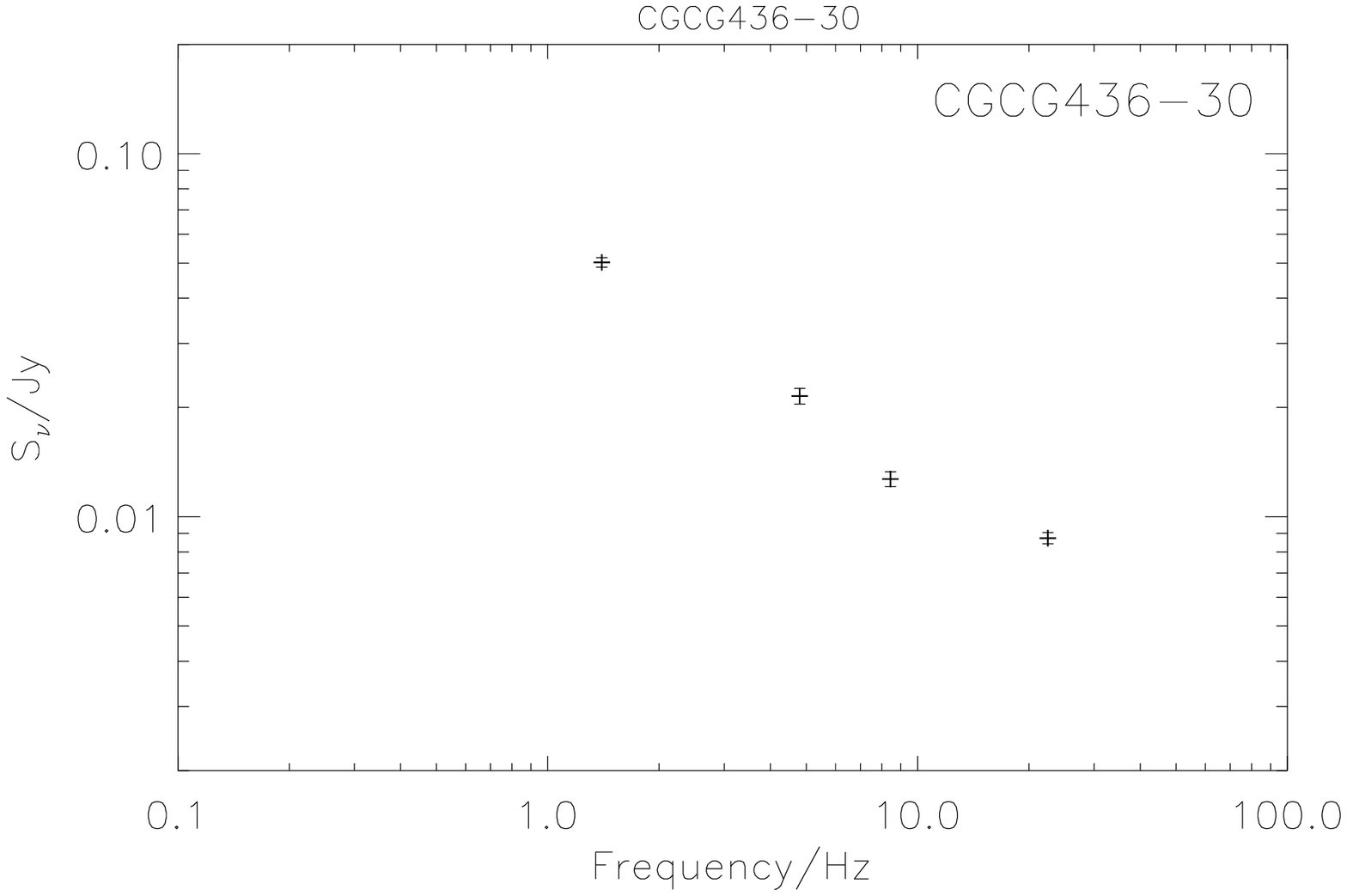}
} \centerline{
\includegraphics[scale = 0.3]{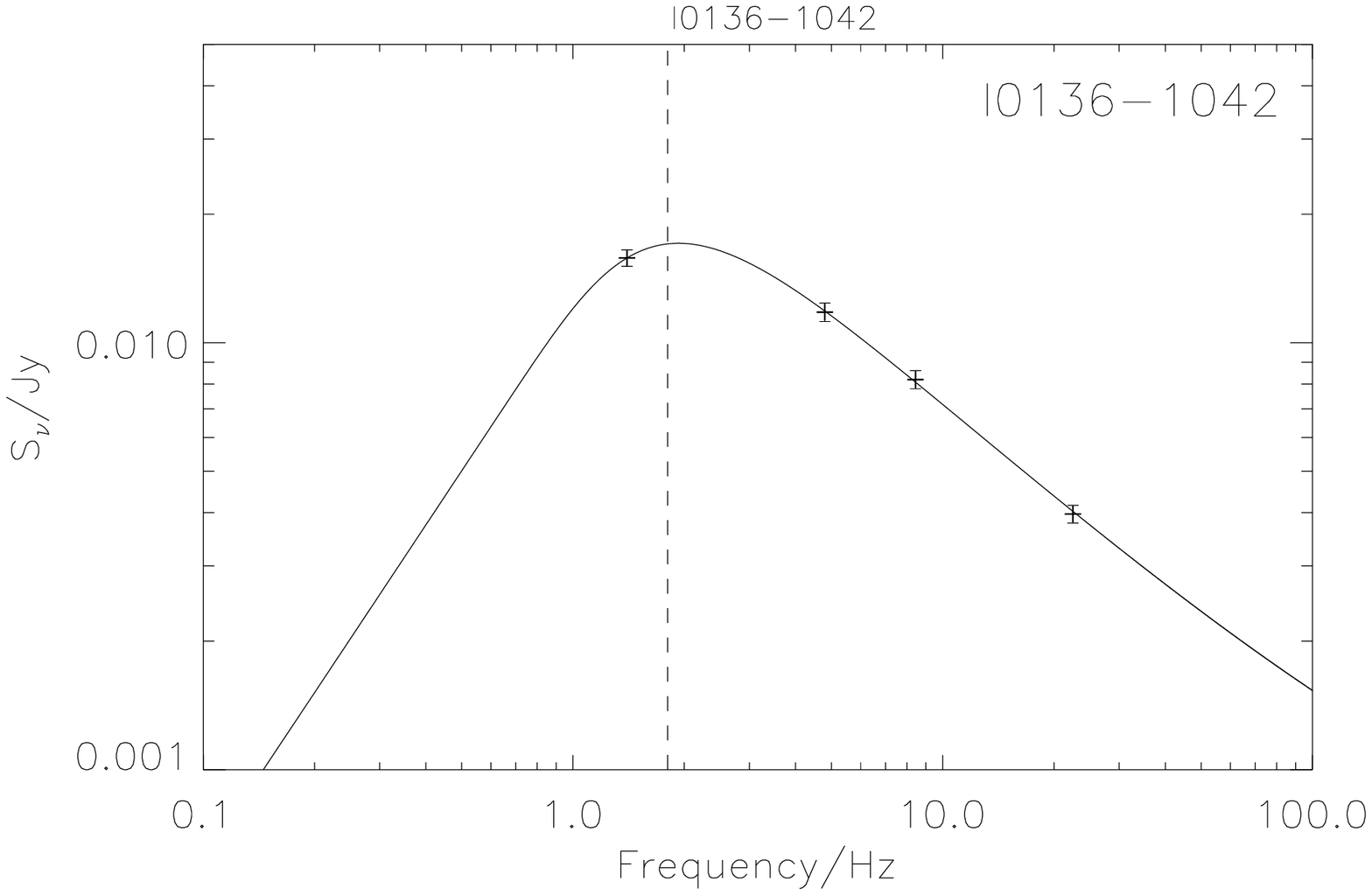}
\includegraphics[scale = 0.3]{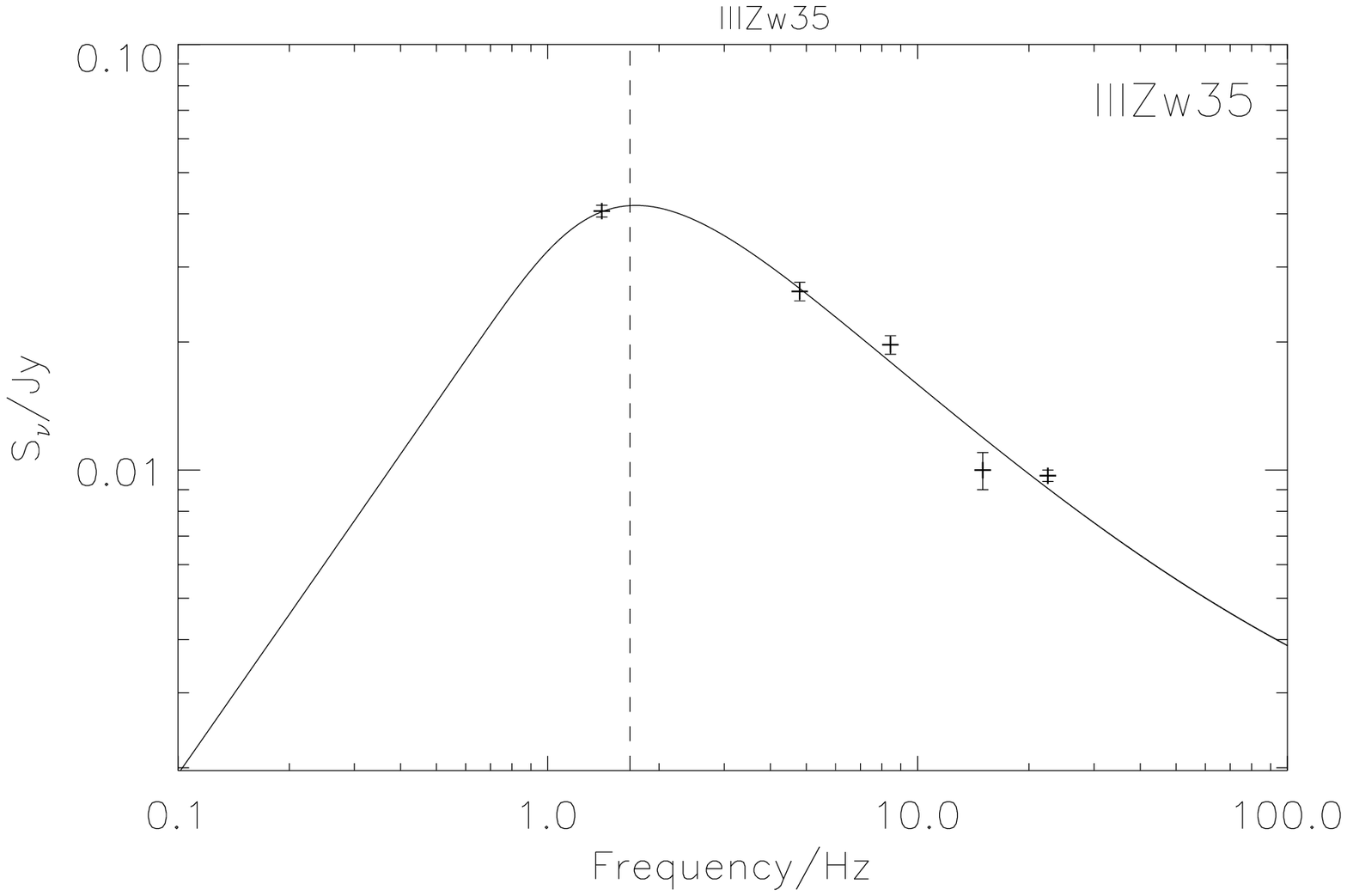}
\includegraphics[scale = 0.3]{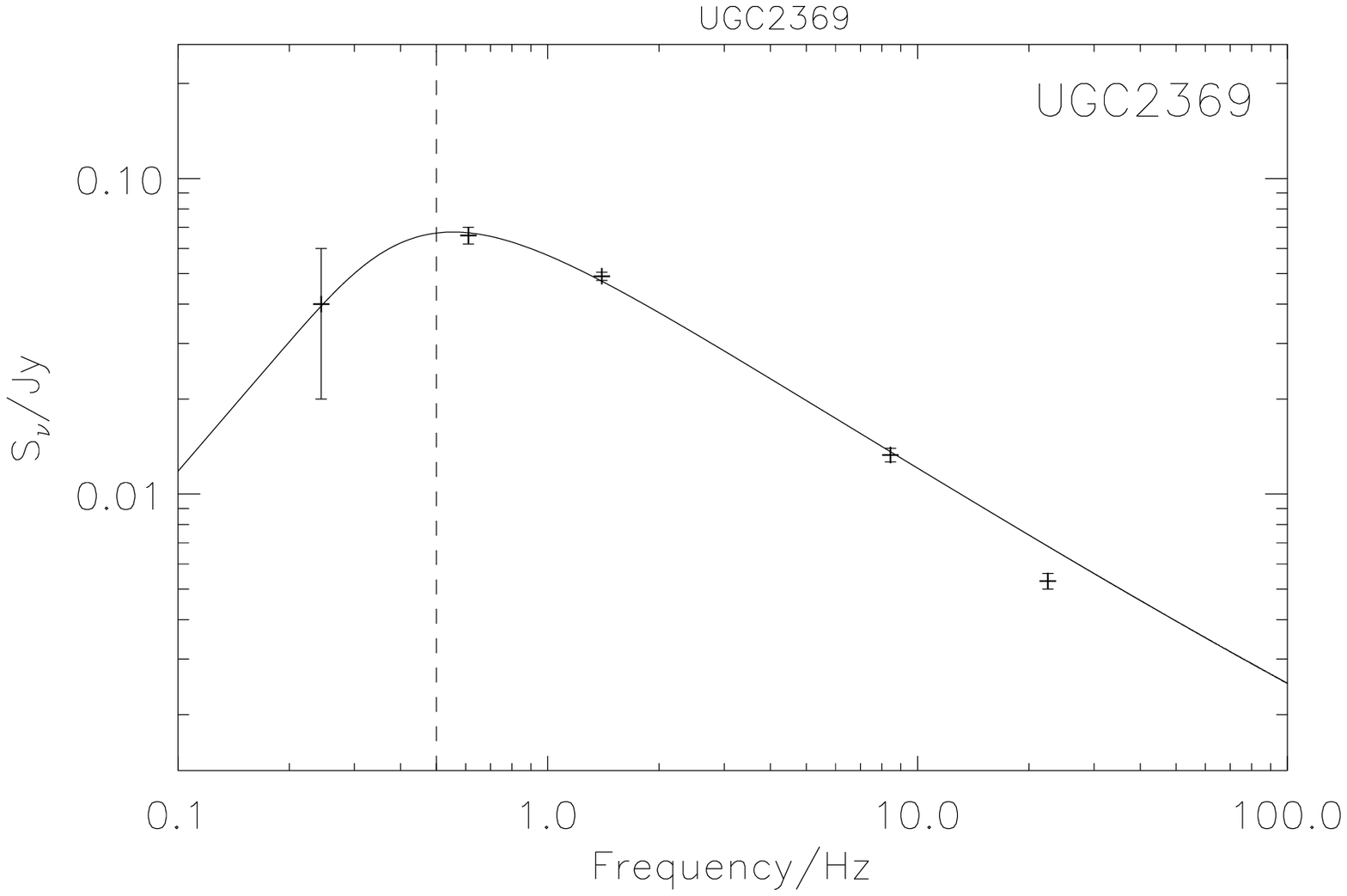}
} \centerline{
\includegraphics[scale = 0.3]{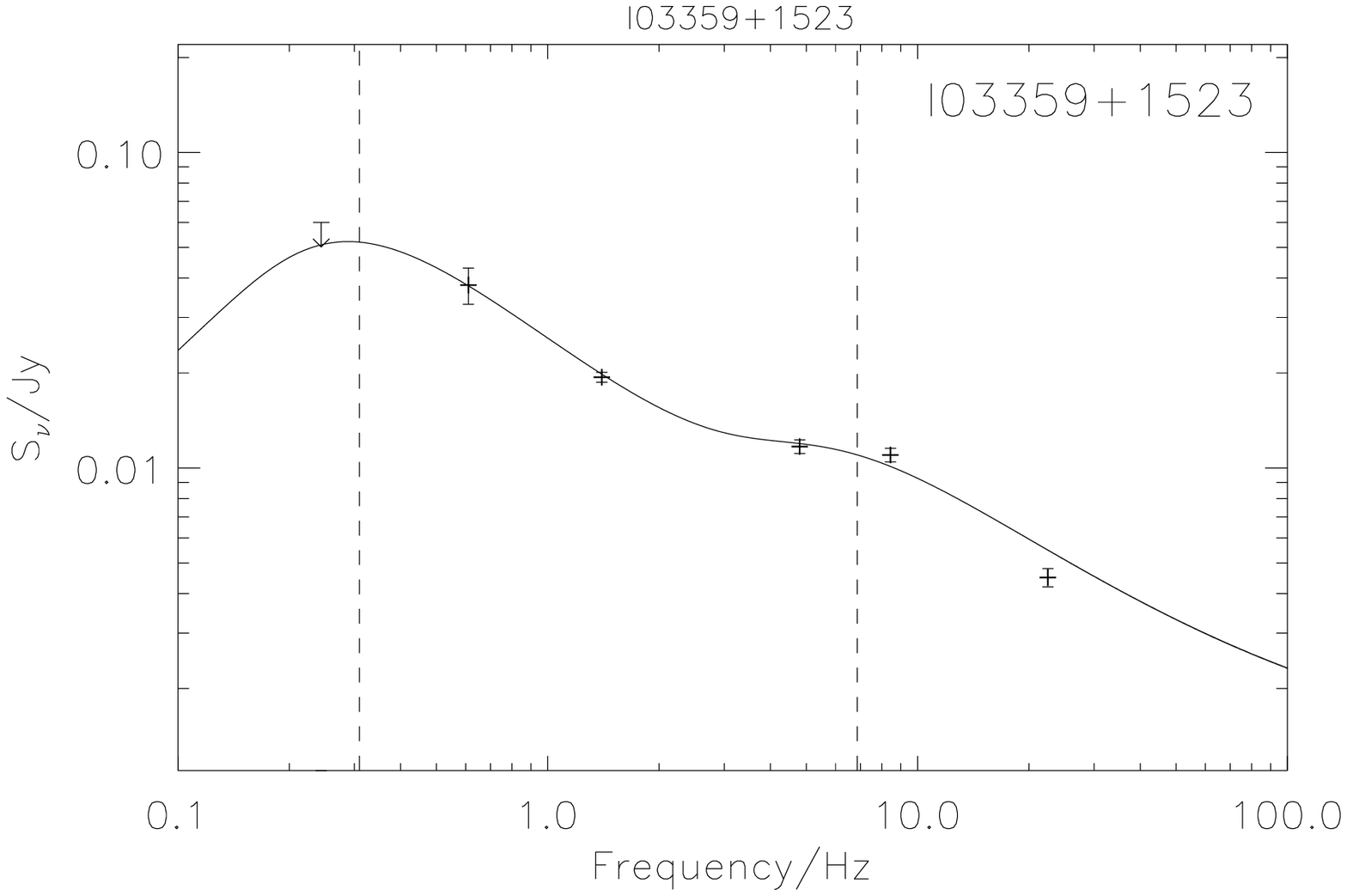}
\includegraphics[scale = 0.3]{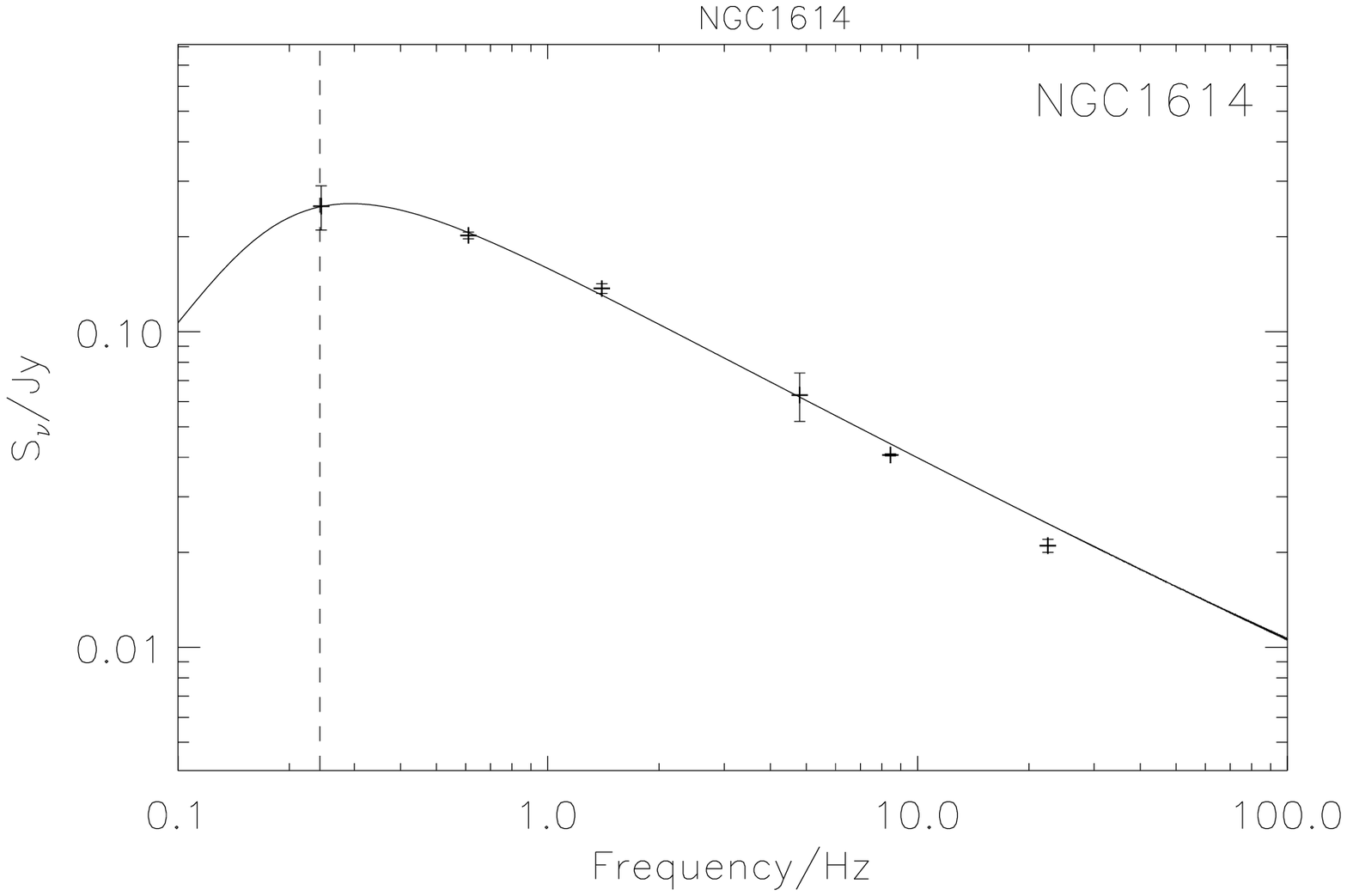}
\includegraphics[scale = 0.3]{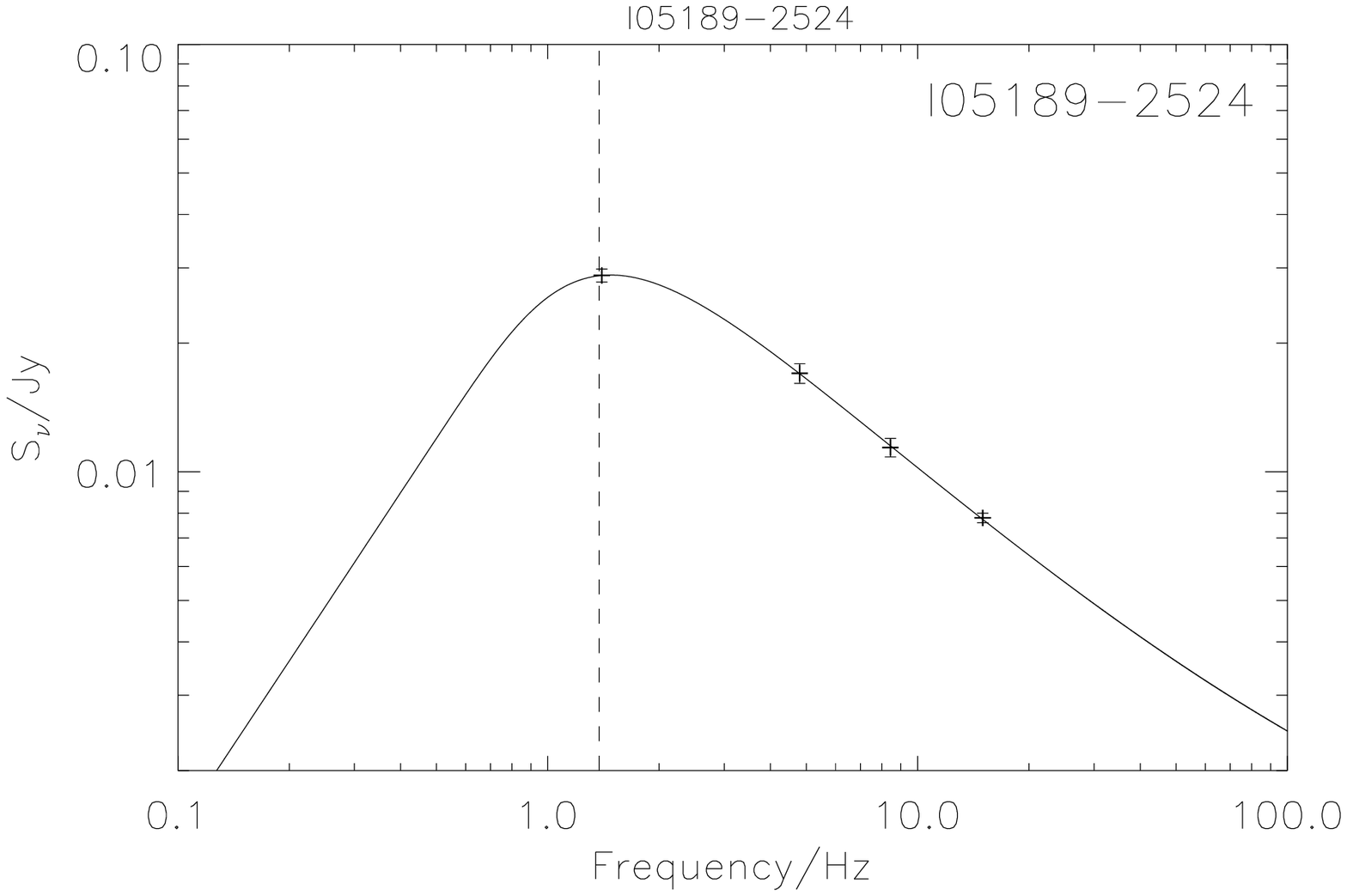}
} \vspace{0.2cm}
\centerline{
\includegraphics[scale = 0.3]{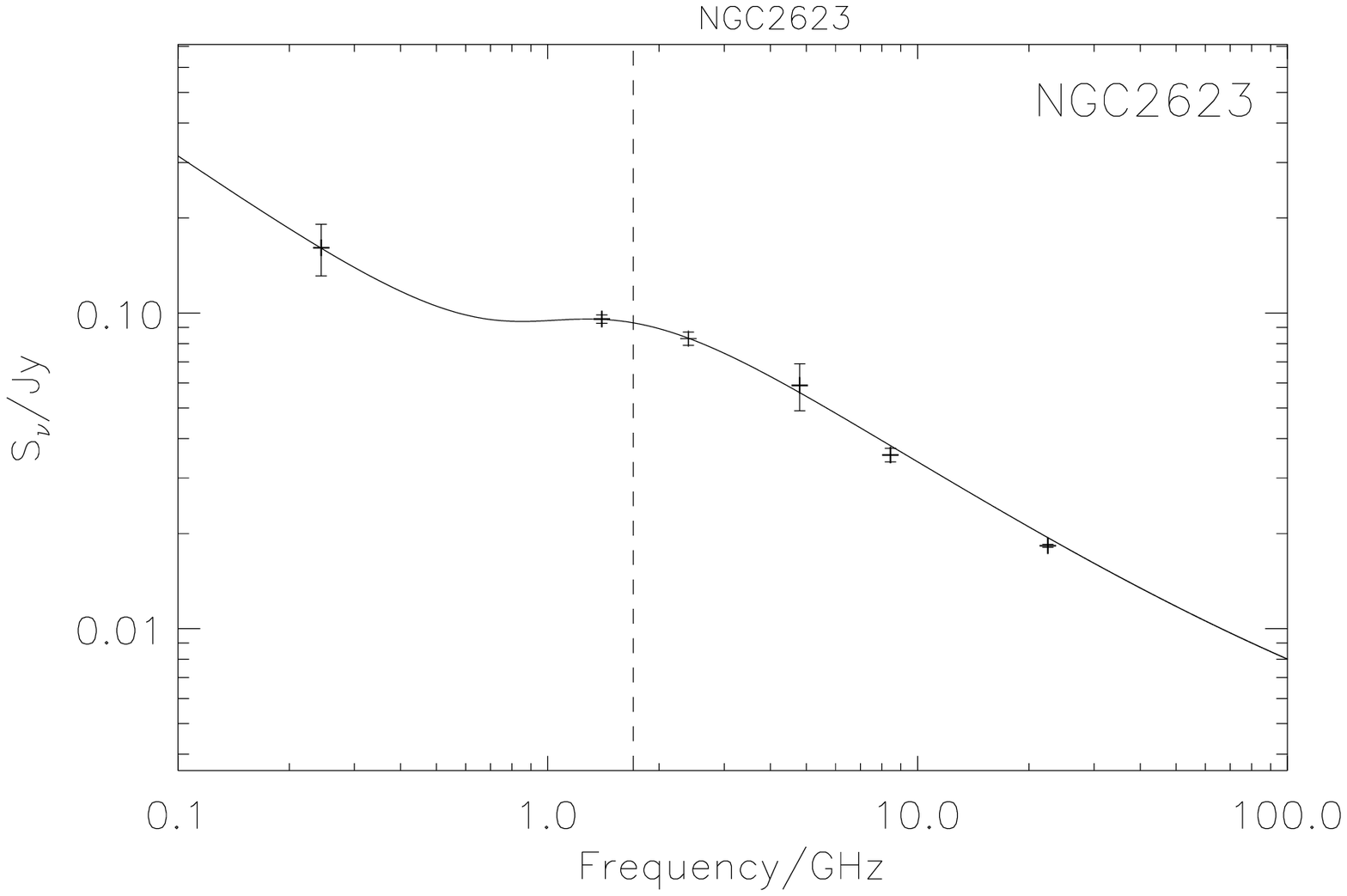}
\includegraphics[scale = 0.3]{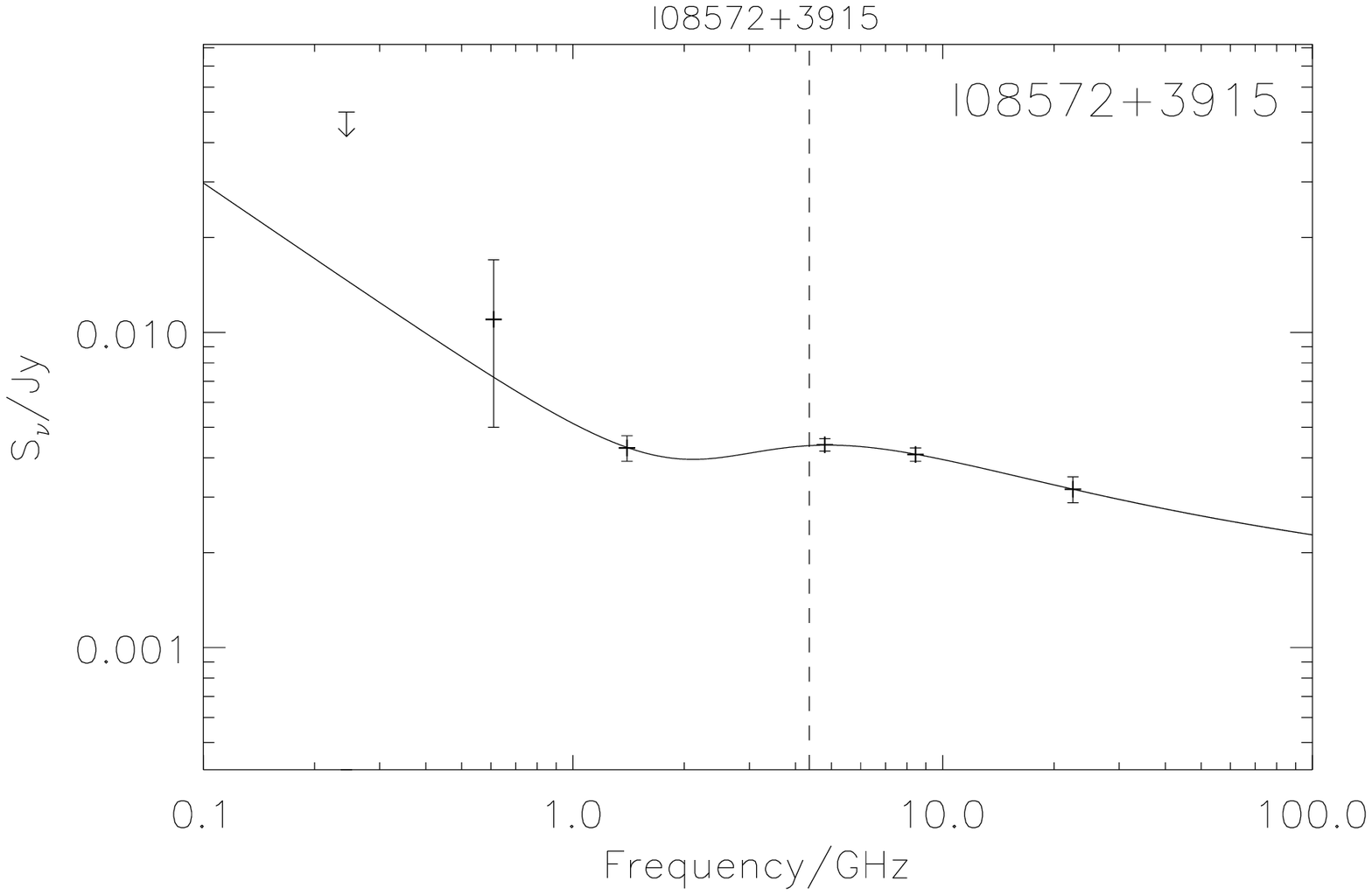}
\includegraphics[scale = 0.3]{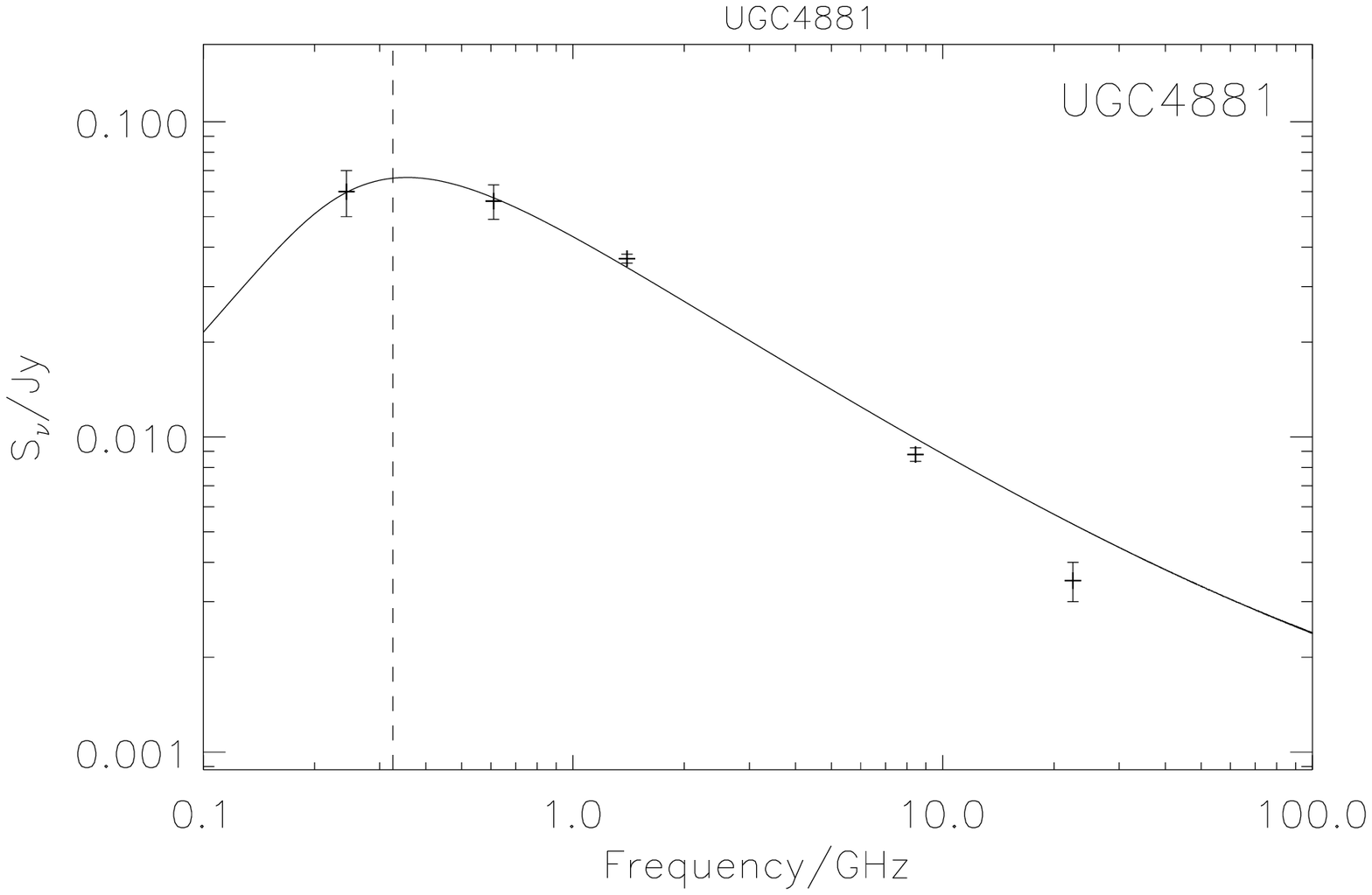}
} 
%\vspace{3cm}
\caption{Model fits to the radio data. The radio fluxes used, other than those given in table~\ref{tab:fluxes}, are collected in Clemens et al. (2008). The turn-over frequencies are indicated with vertical lines. 
Poor quality fits in which the adopted model seems insufficient are shown as dot-dashed lines.} 
\end{figure*}

\begin{figure*}
\addtocounter{figure}{-1}
%\vspace{-3cm}   
\centerline{
\includegraphics[scale = 0.3]{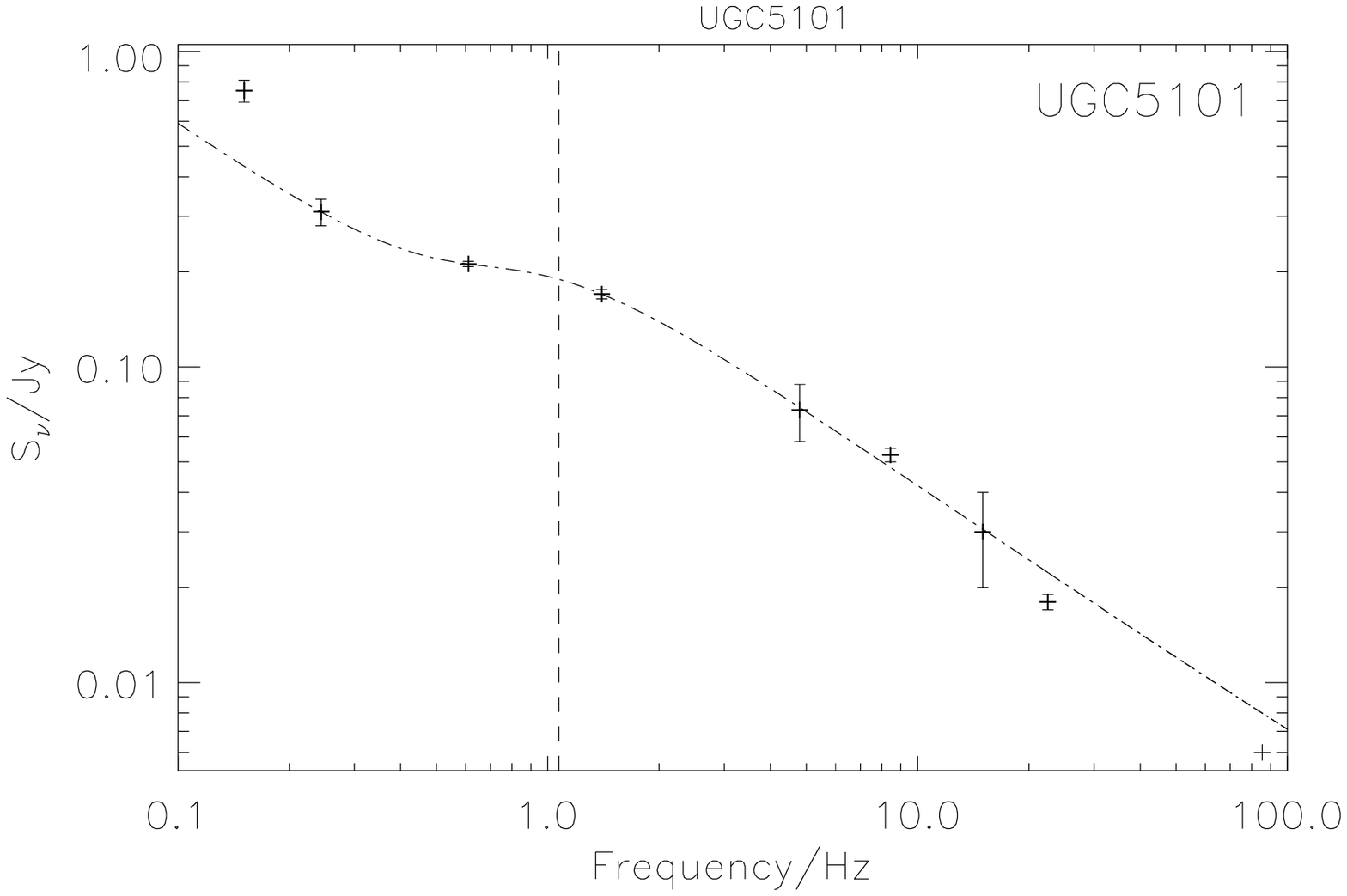}
\includegraphics[scale = 0.3]{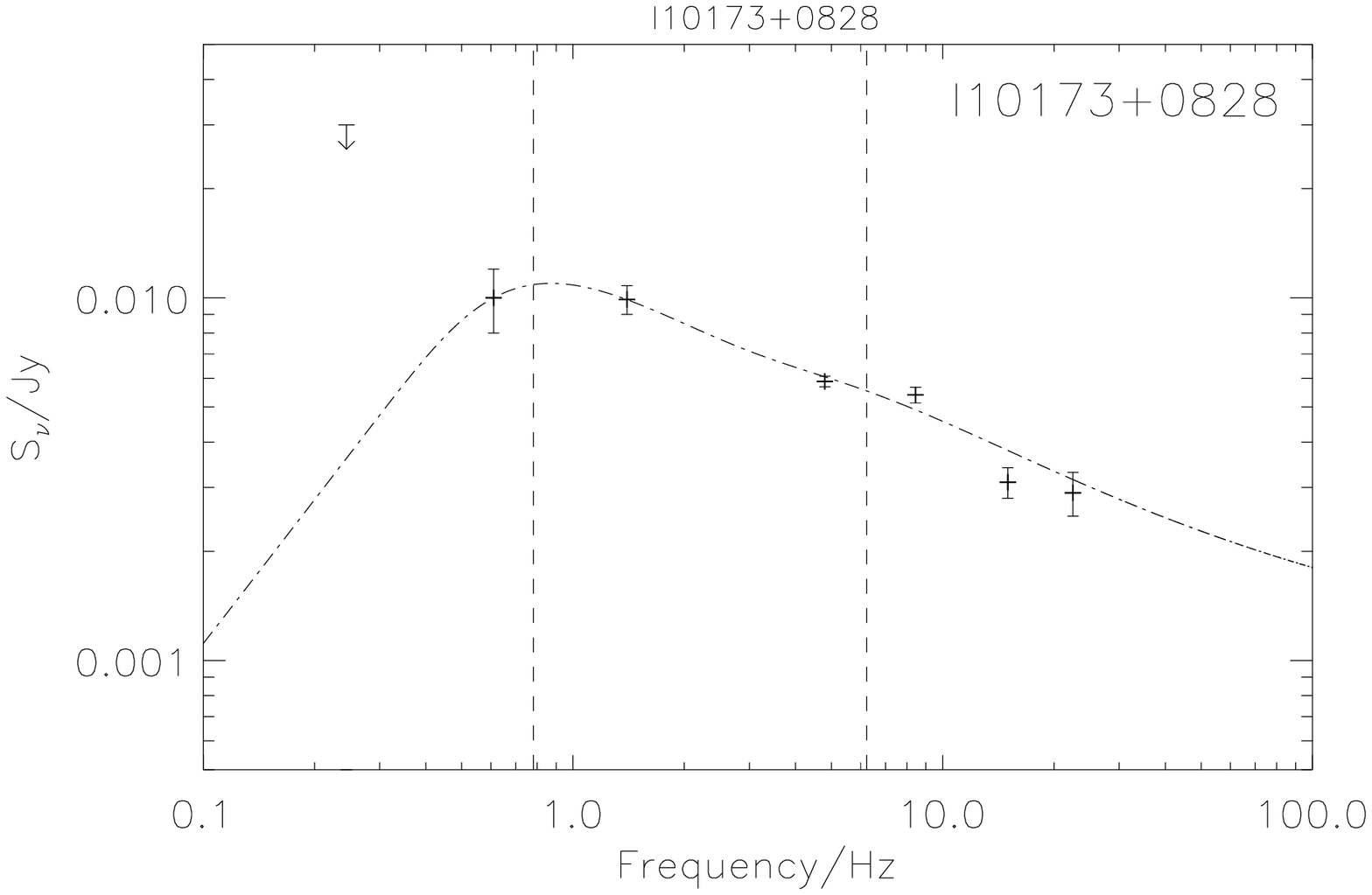}
\includegraphics[scale = 0.3]{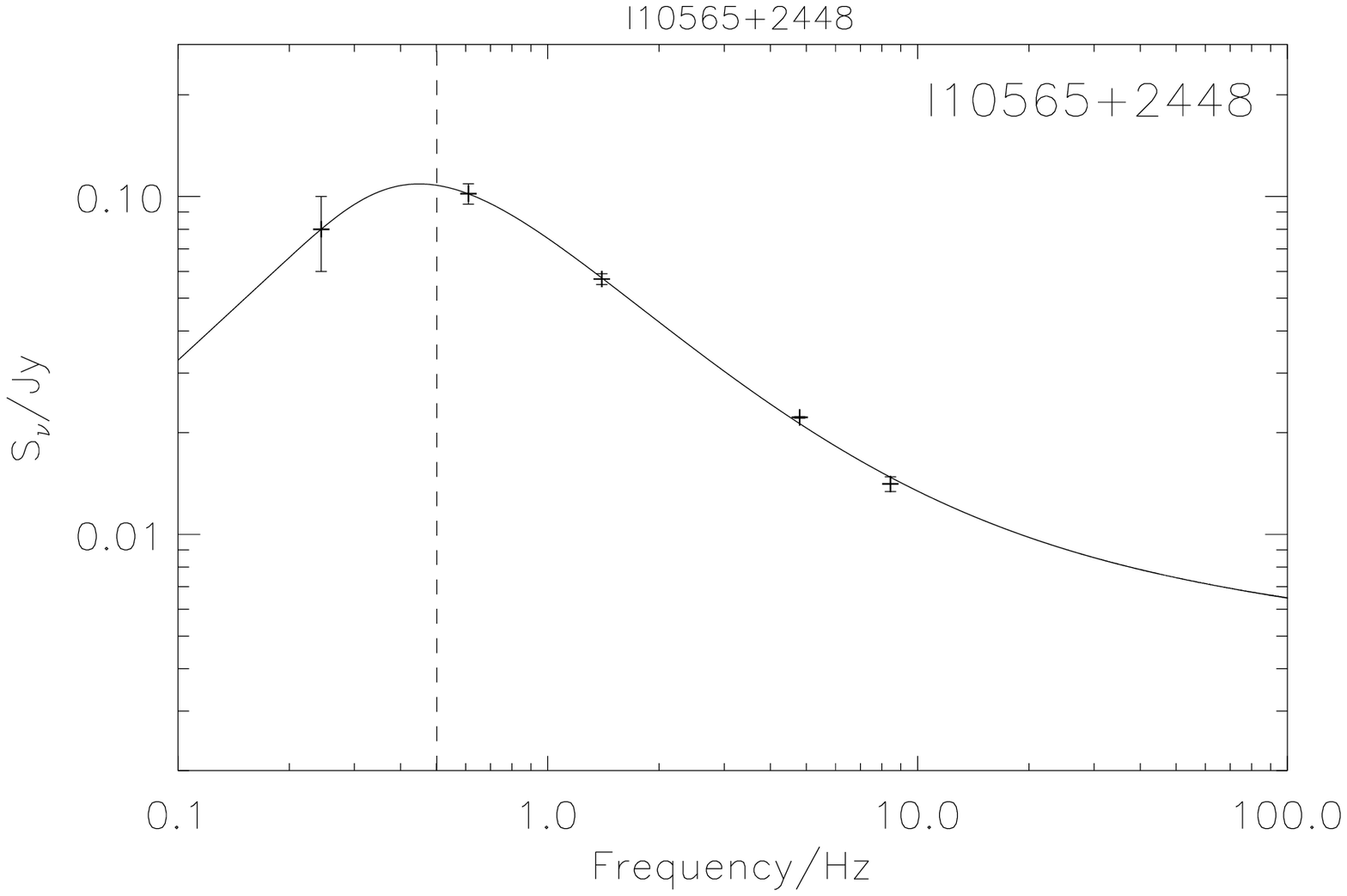}
} \centerline{
\includegraphics[scale = 0.3]{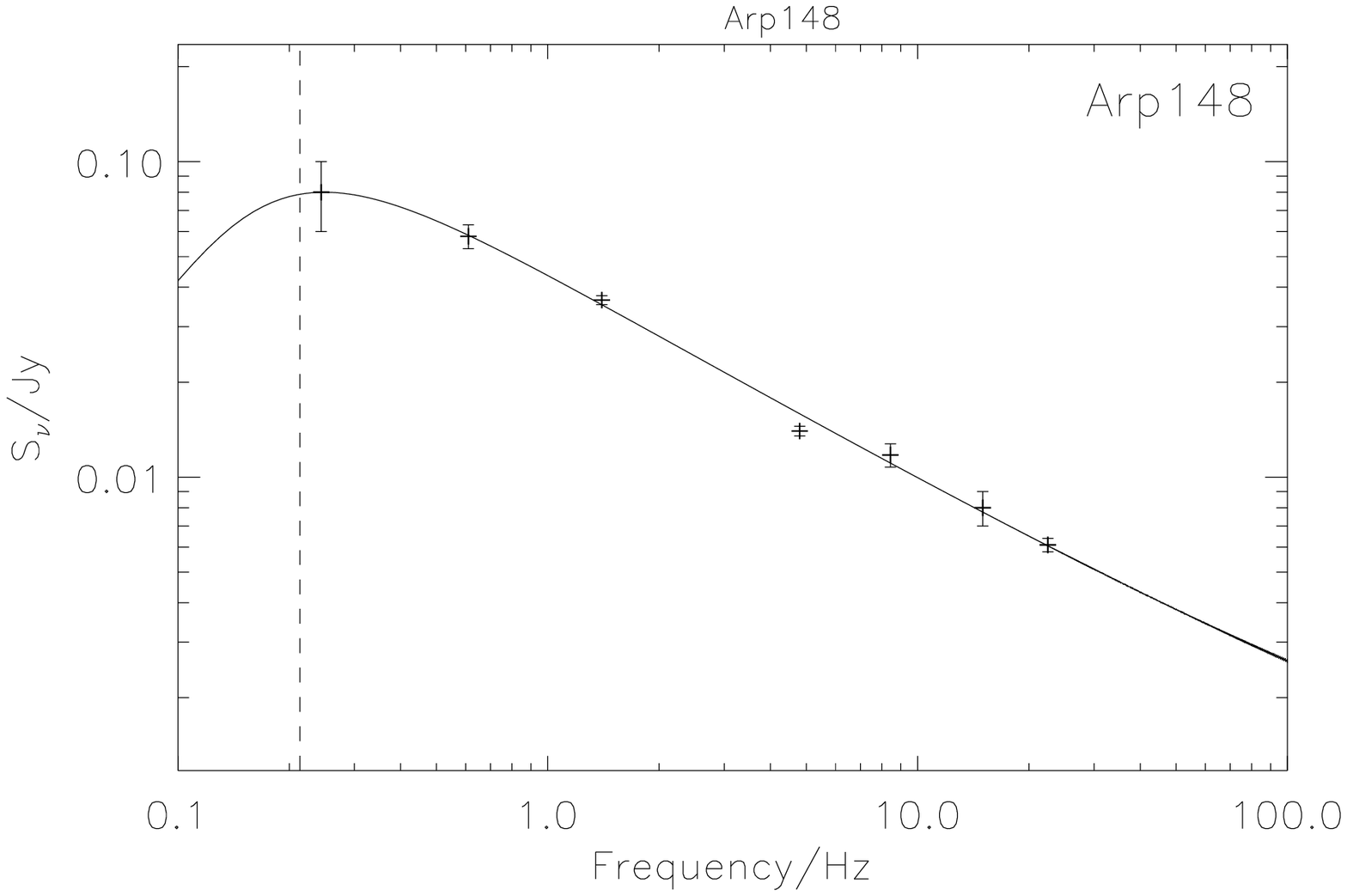}
\includegraphics[scale = 0.3]{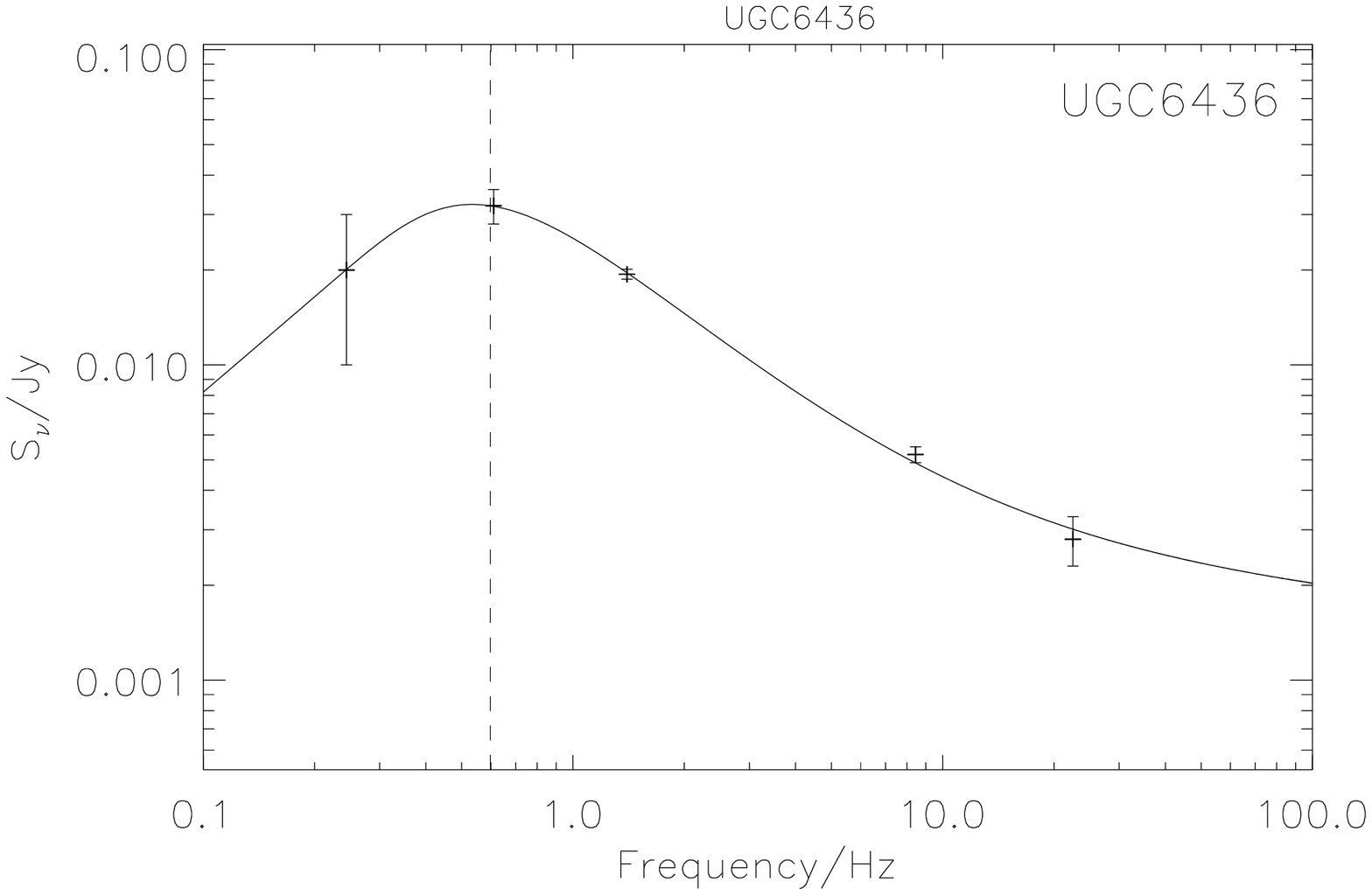}
\includegraphics[scale = 0.3]{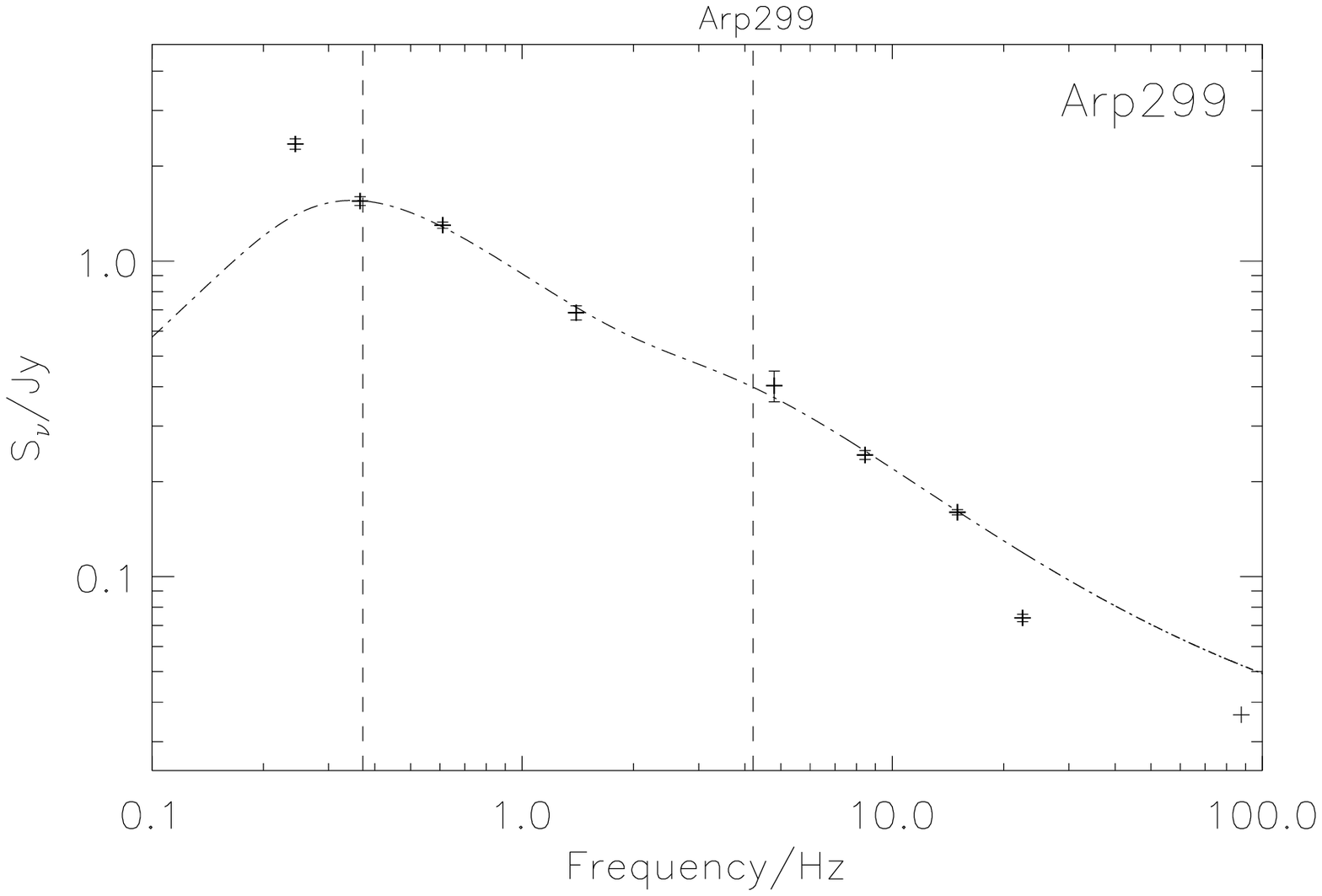}
} \centerline{
\includegraphics[scale = 0.3]{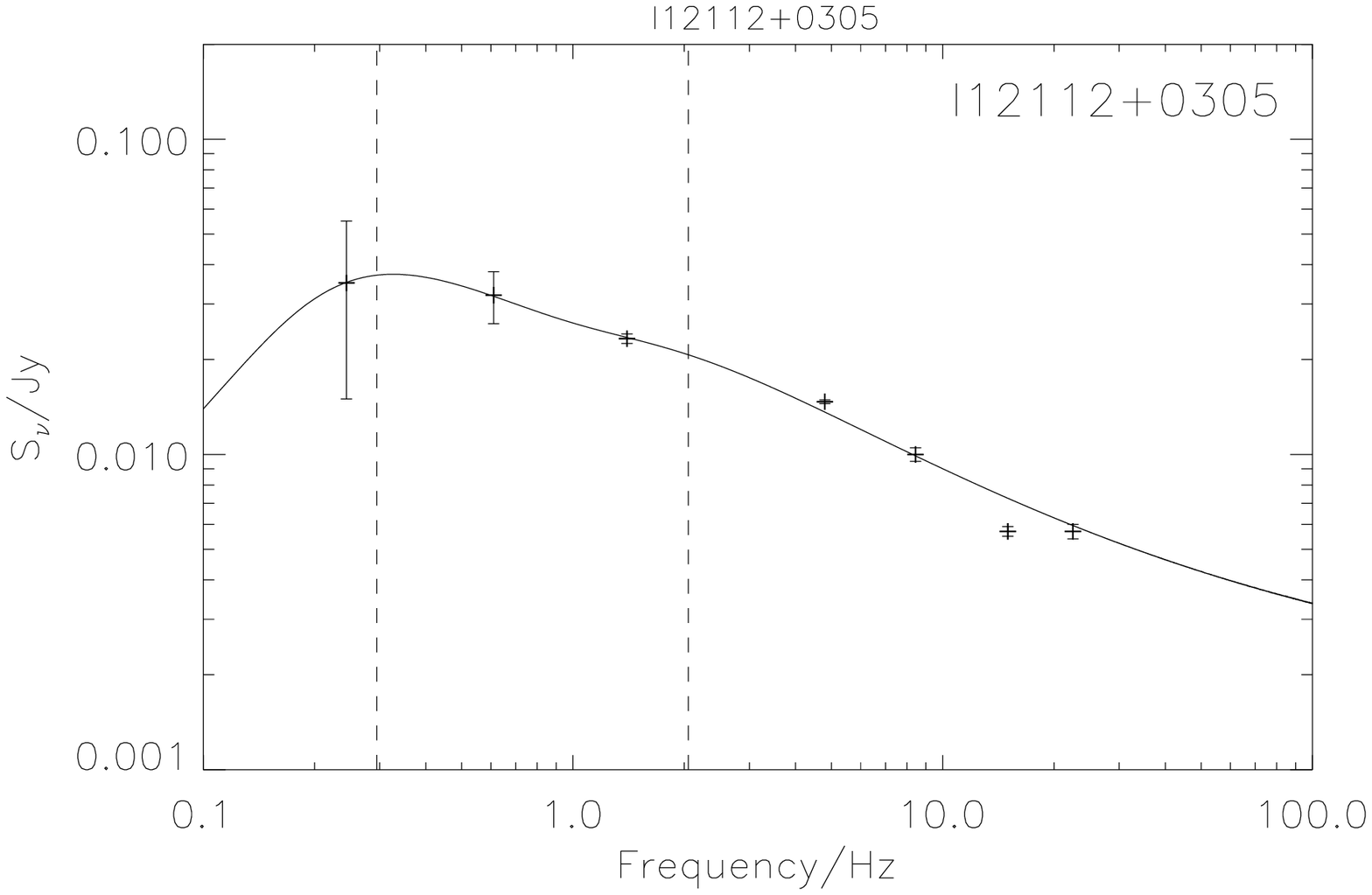}
\includegraphics[scale = 0.3]{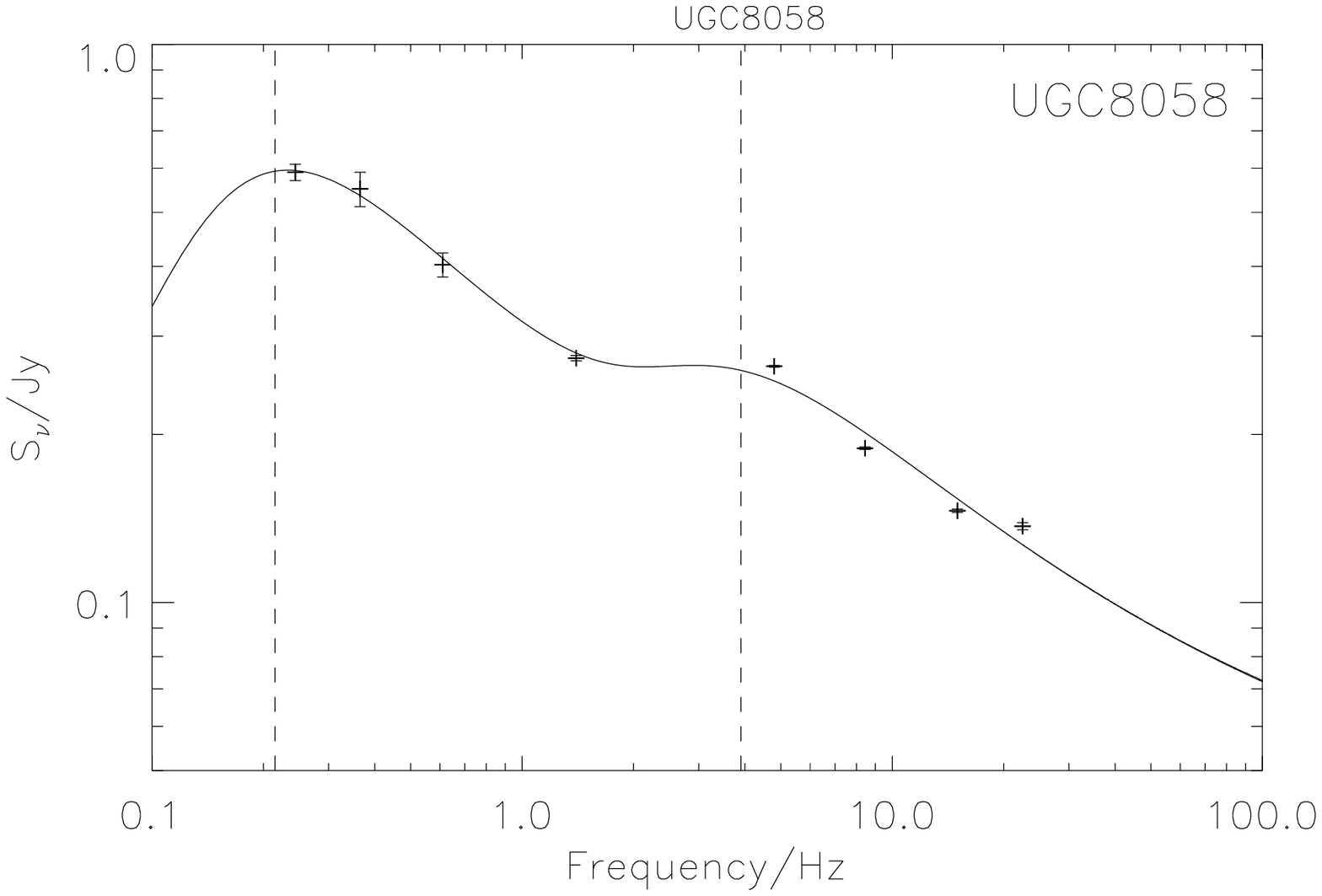}
\includegraphics[scale = 0.3]{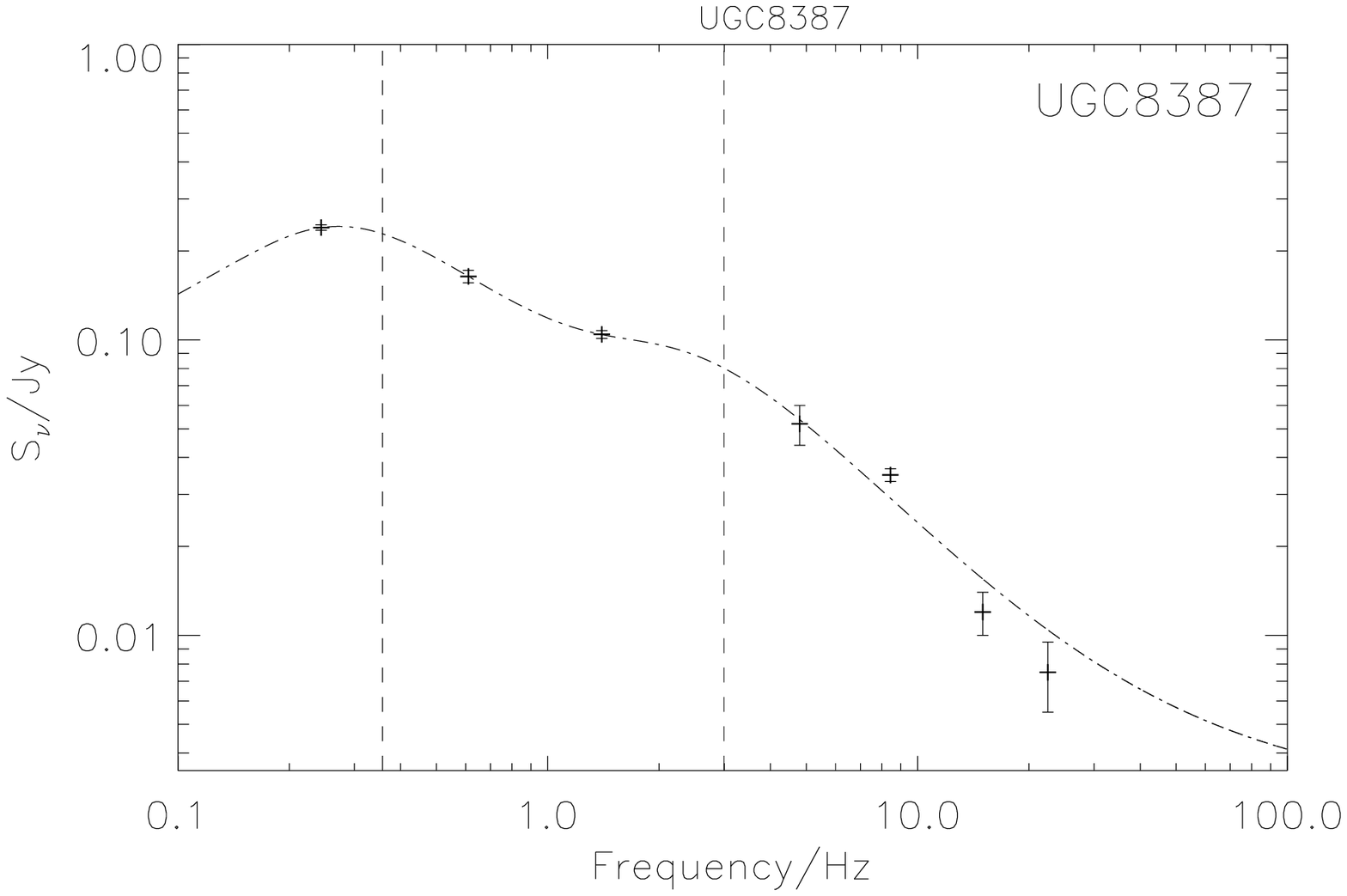}
} \vspace{0.2cm}
\centerline{
\includegraphics[scale = 0.3]{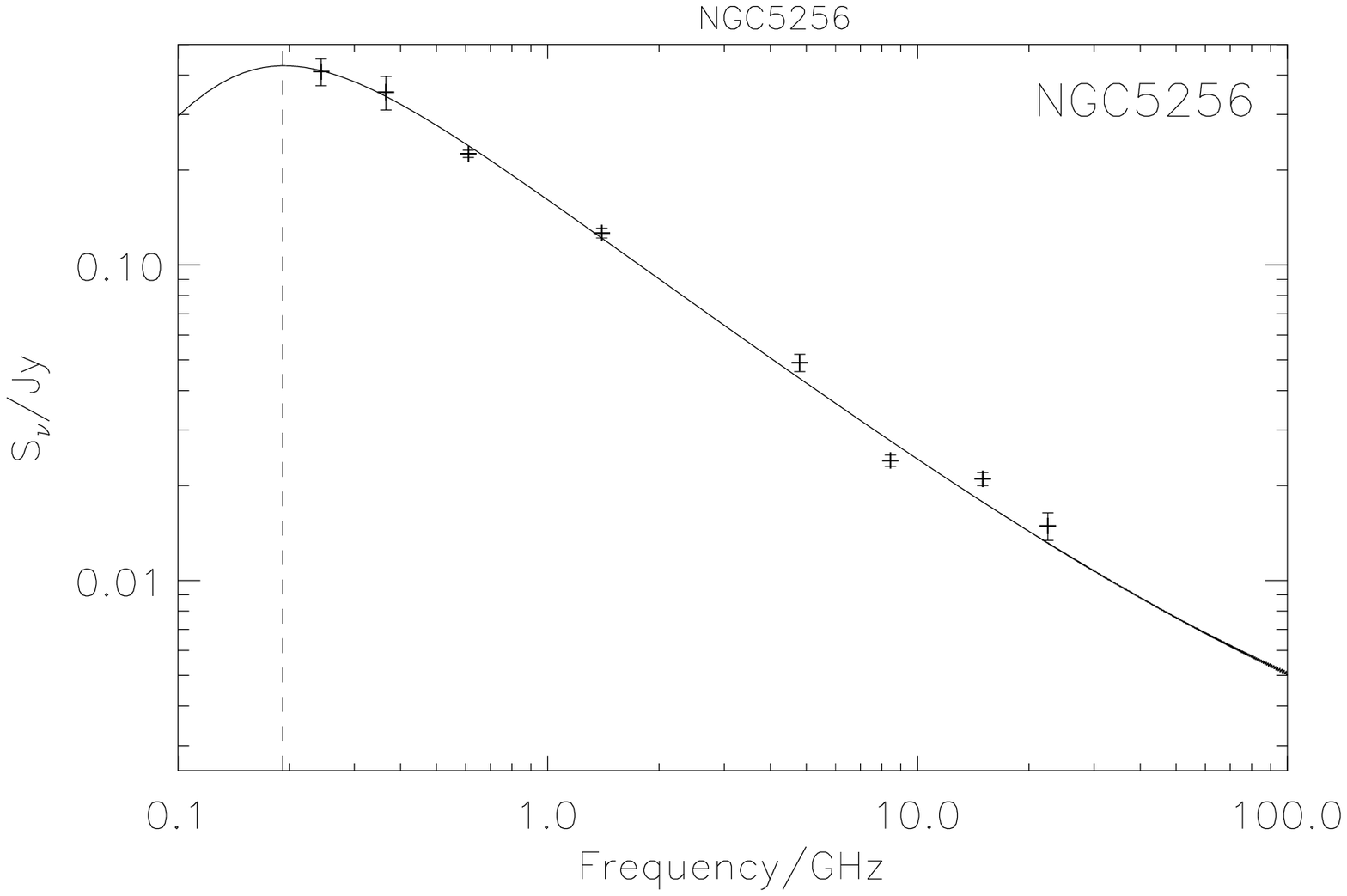}
\includegraphics[scale = 0.3]{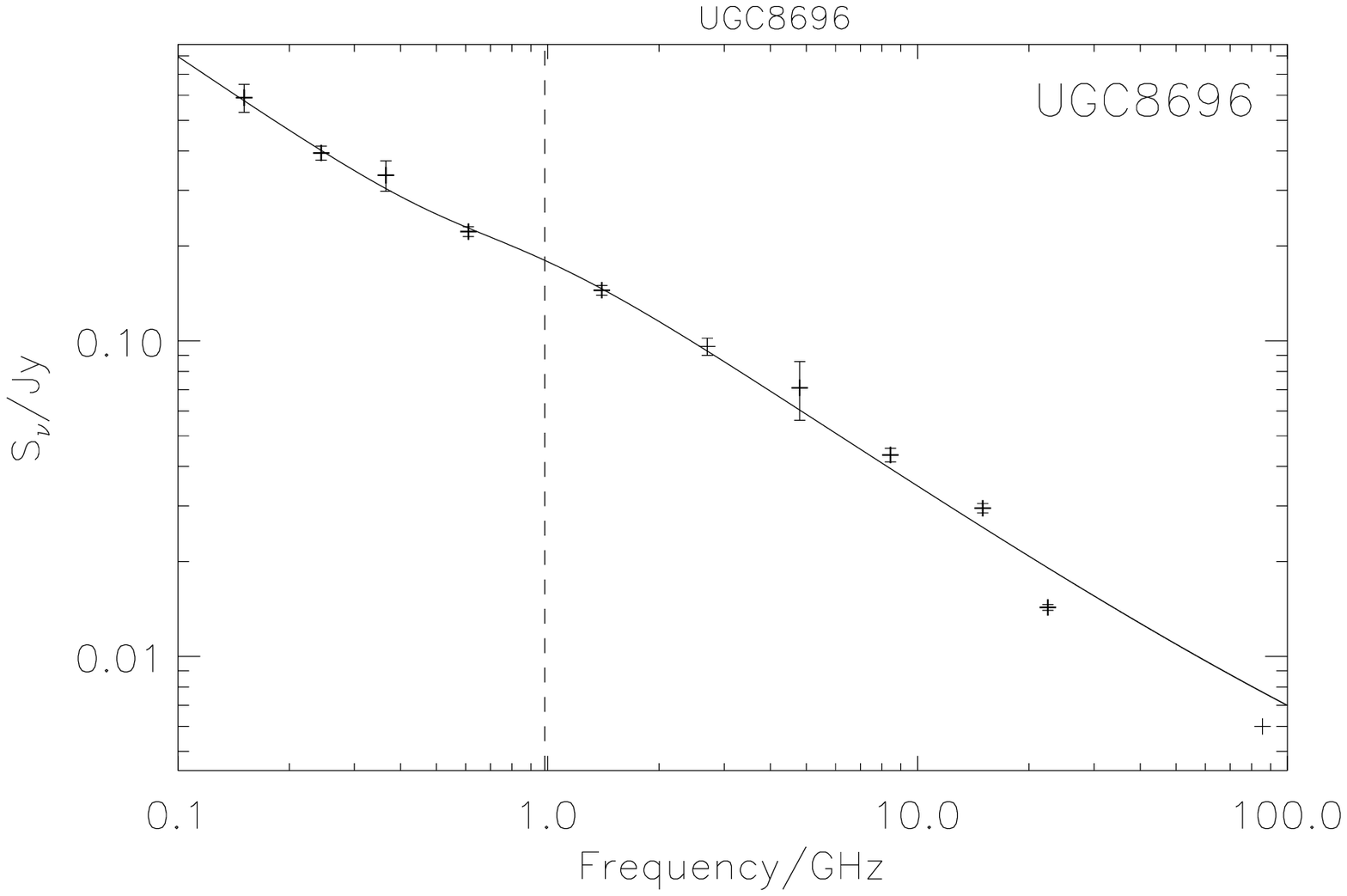}
\includegraphics[scale = 0.3]{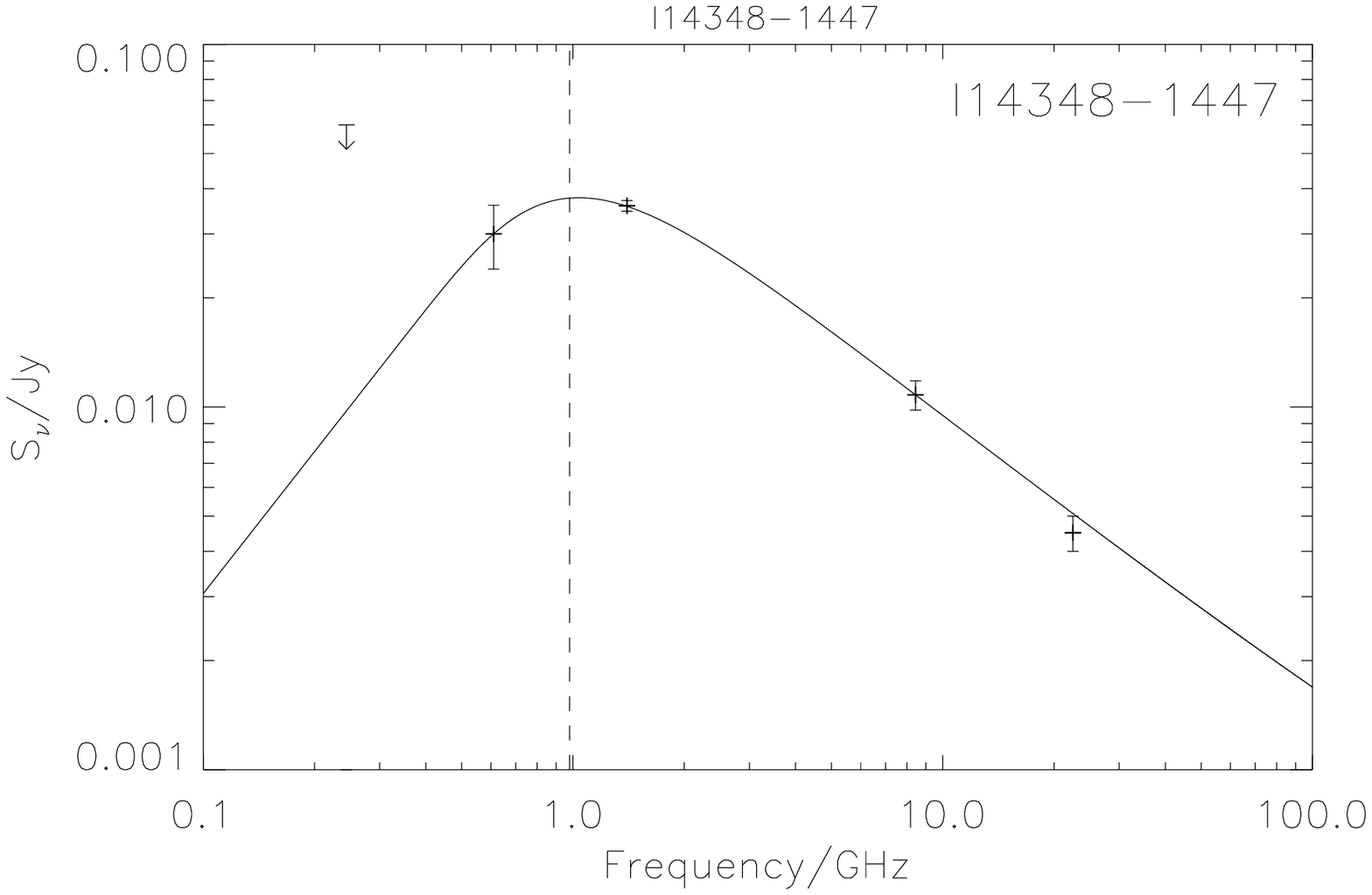}
} 
%\vspace{3cm}
\caption{Continued} 
\end{figure*}

\begin{figure*}
\addtocounter{figure}{-1}
%\vspace{-3cm}   
\centerline{
\includegraphics[scale = 0.3]{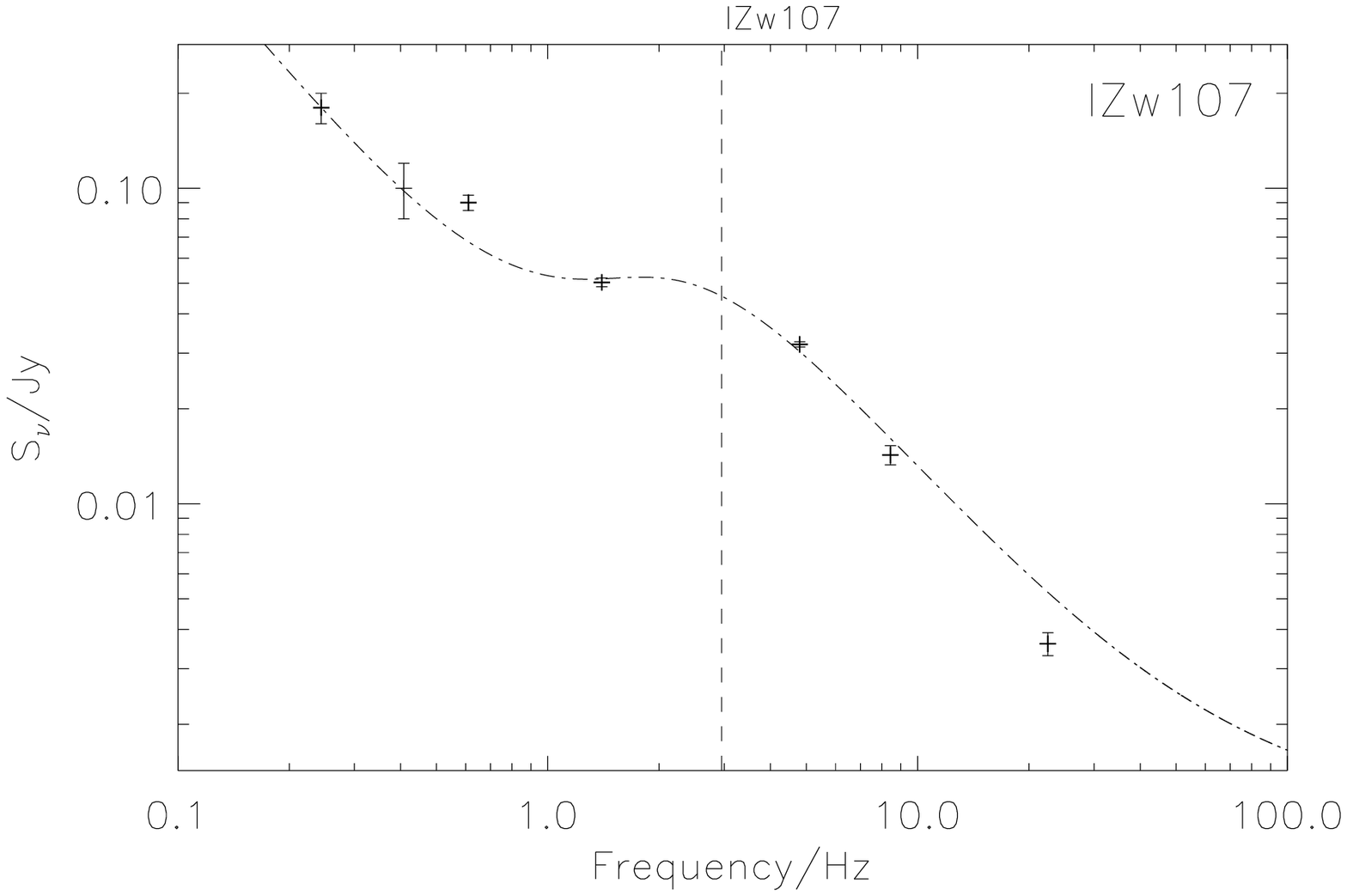}
\includegraphics[scale = 0.3]{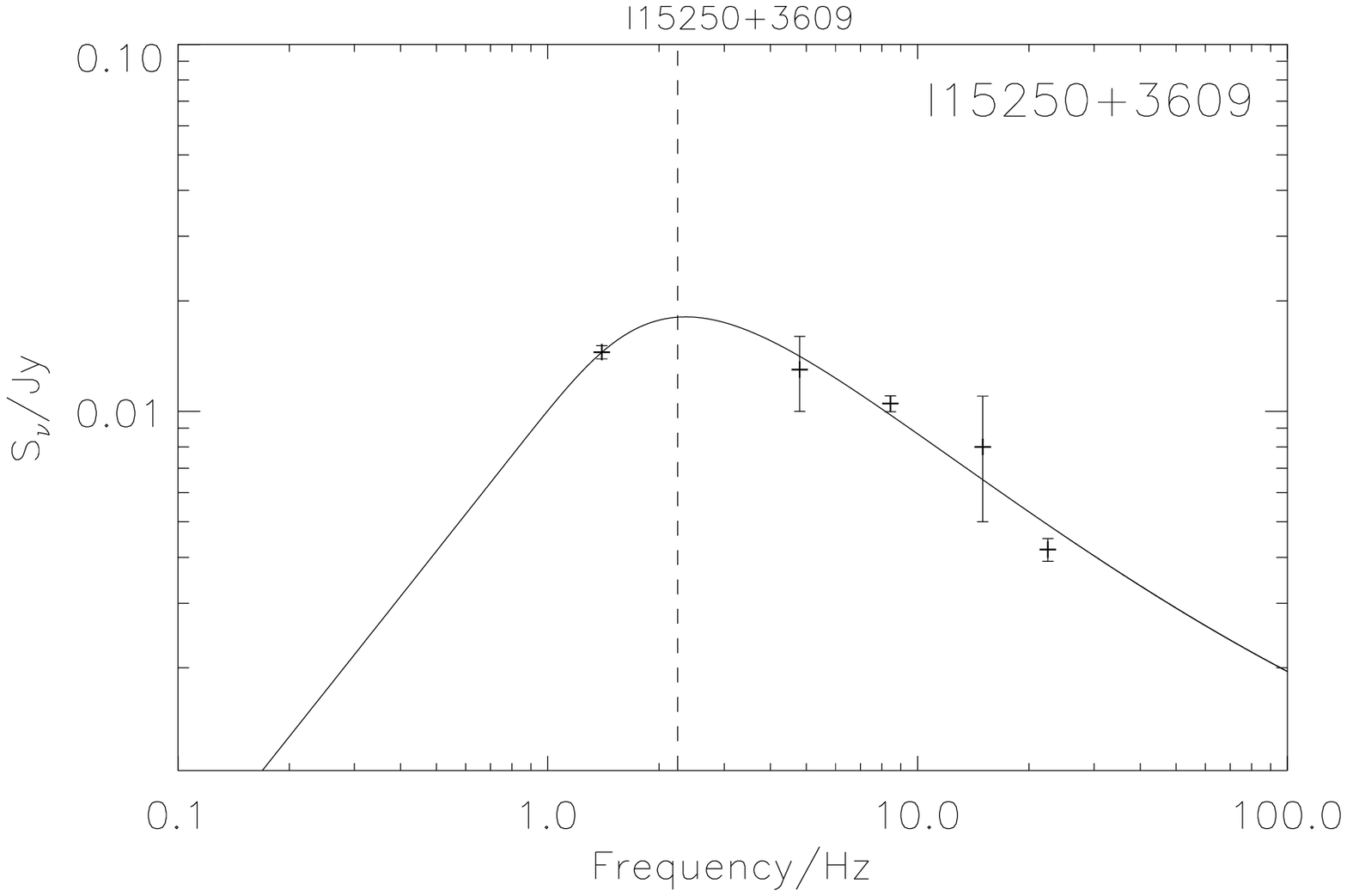}
\includegraphics[scale = 0.3]{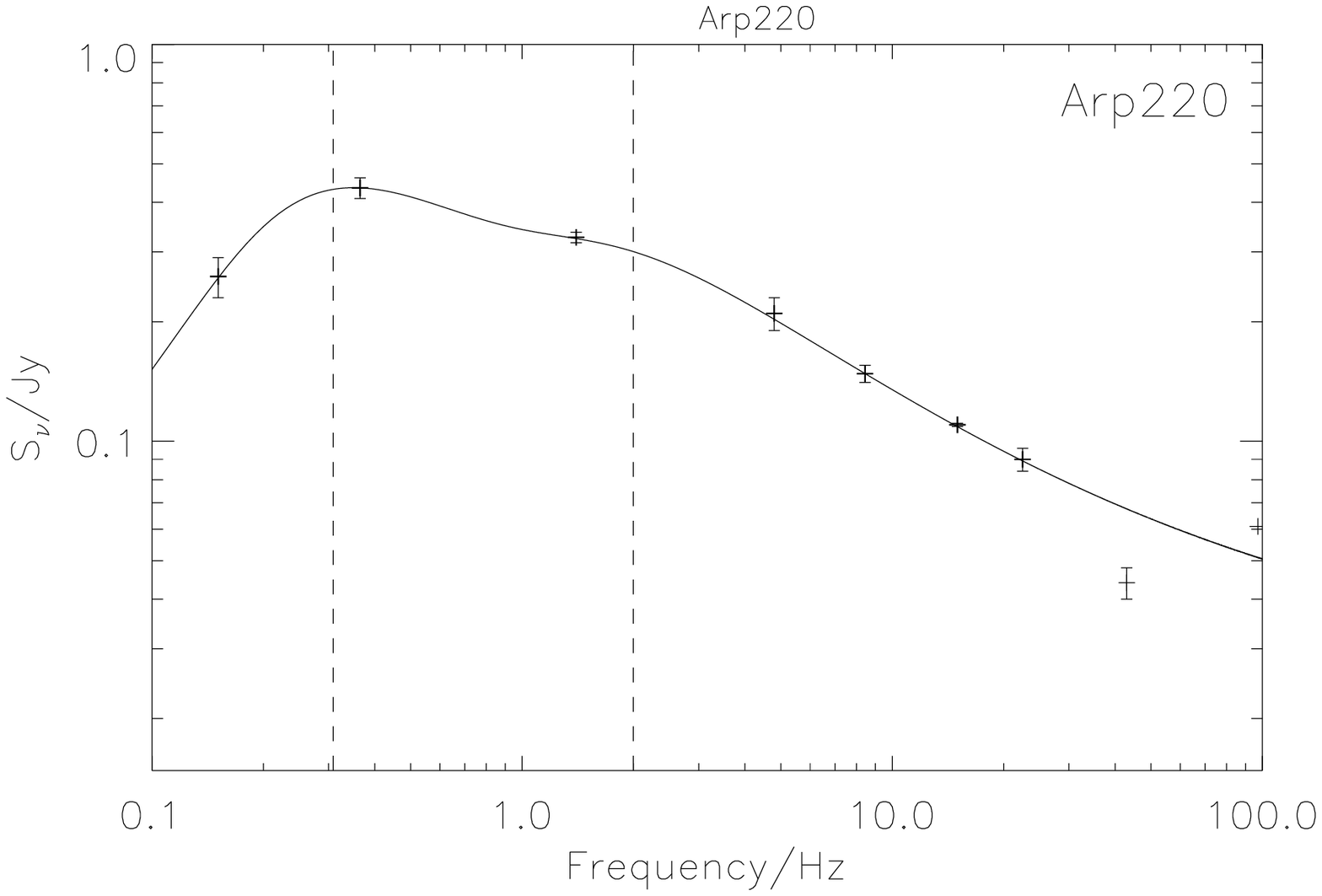}
} \centerline{
\includegraphics[scale = 0.3]{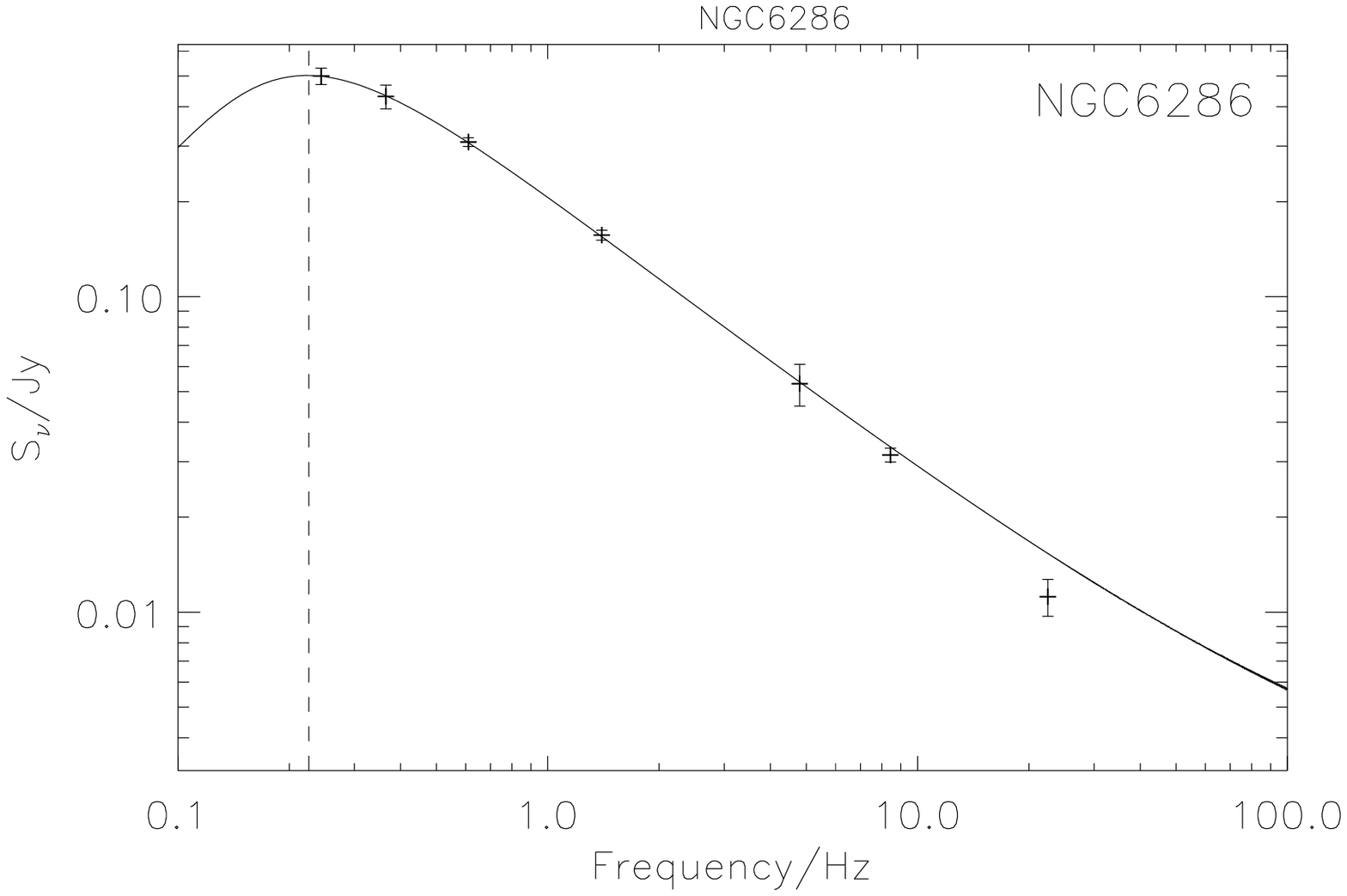}
\includegraphics[scale = 0.3]{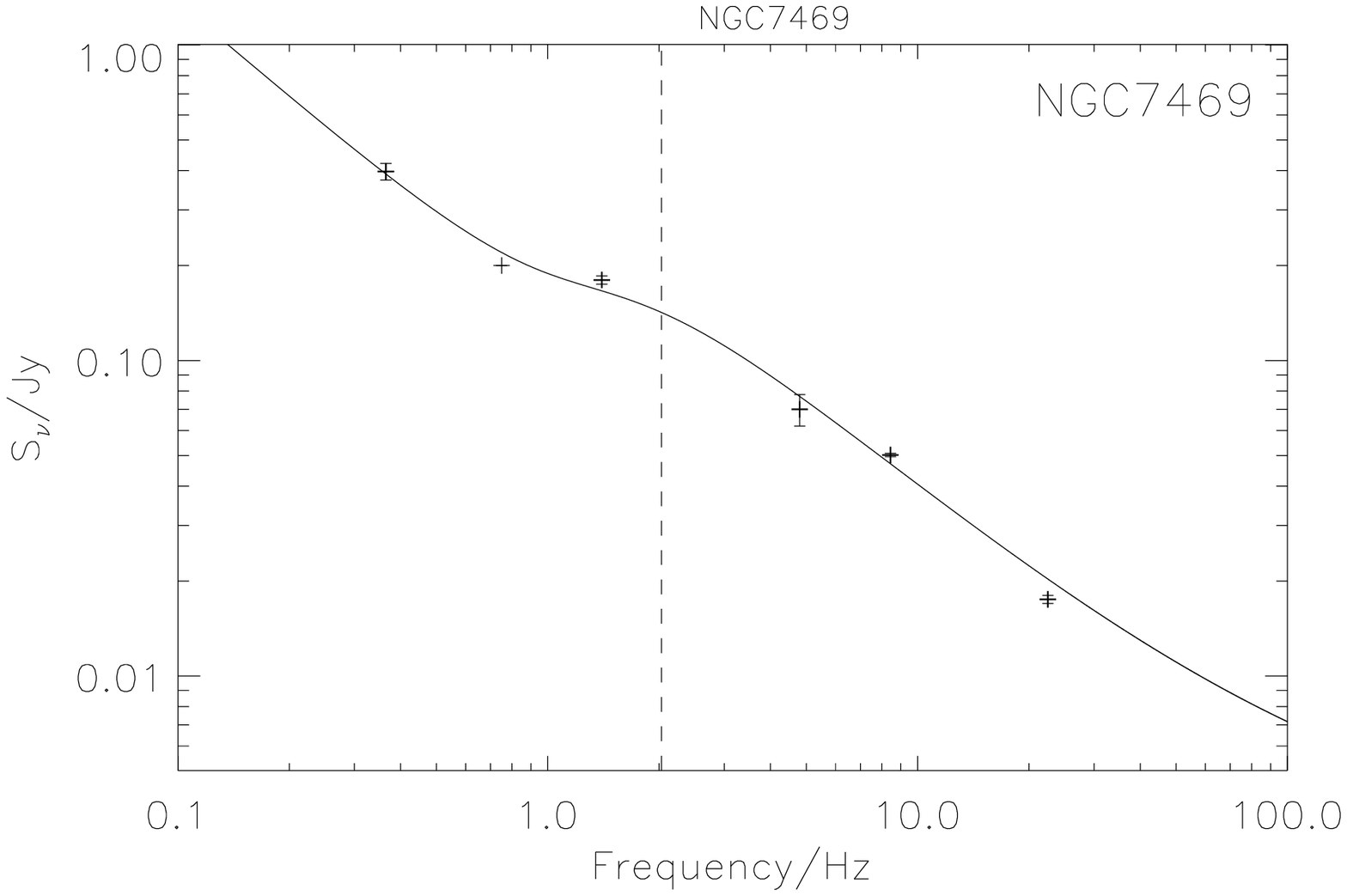}
\includegraphics[scale = 0.3]{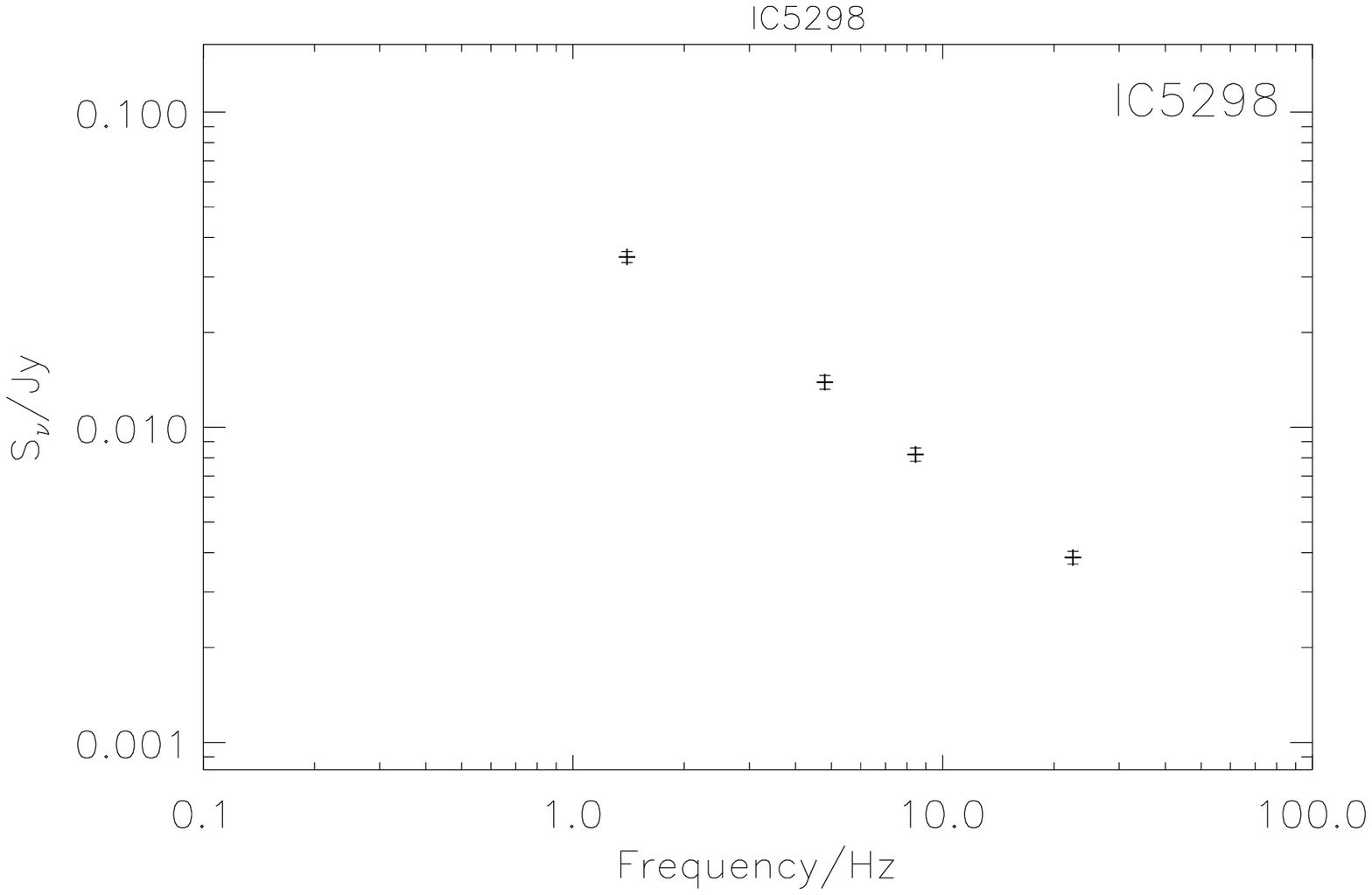}
} \vspace{0.2cm}
\centerline{
\includegraphics[scale = 0.3]{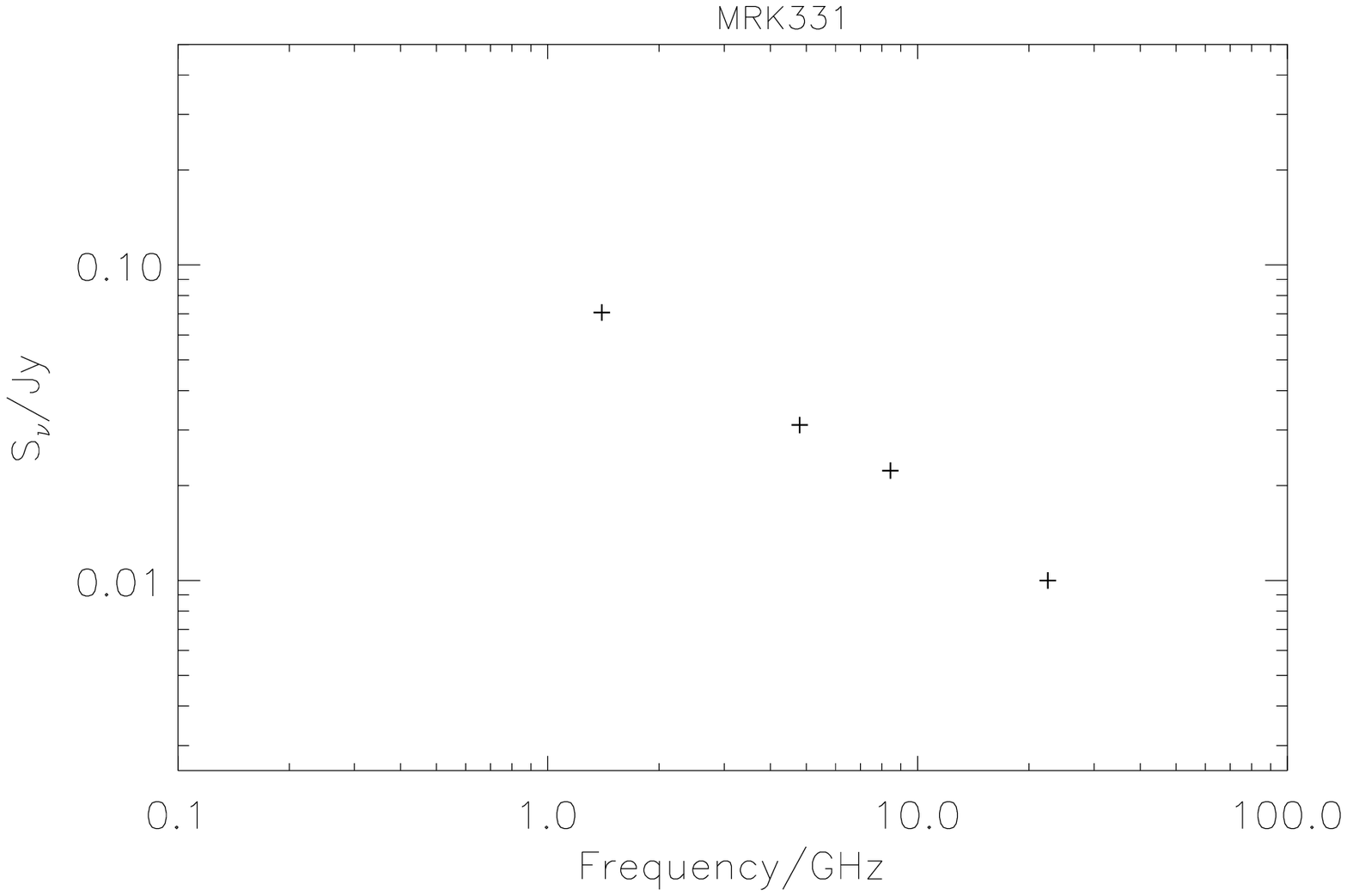}
}
%\vspace{3cm} 
\caption{Continued} 
\label{fig:fits}
\end{figure*}

\begin{table*}
\caption{Fitted model parameters. Values that are very uncertain are shown in parentheses. Derived star 
formation rates refer to the mass range 0.15-120$\;\rm M_{\odot}$. $^*$ parameter held fixed at this value. 
$^{\dagger}$ UGC~8058 contains a radio loud AGN that emits above 4.8~GHz, see section~\ref{sec:UGC8058}.}
\centering
\begin{tabular}{c c c c c c c c c c}
\hline\hline
Source          &model& $\alpha$ &$H$ & $\nu_{t1}$ & $\nu_{t2}$ & $f_2/f_1(8.4)$ & $L_{8.4}$   & SFR & $x$\\
                &     &  & & GHz        & GHz        &                & $\rm 10^{22}\,W\,Hz^{-1}$ & $\rm M_{\odot}yr^{-1}$ & \\
\hline
%                 mod      alp     H       nu_1        nu_2   f2/f1     L8.4   SFR        x 
NGC~34          & --- &         &         & $\lesssim$1 & ---  & ---    & 1.08  & 28      &  ---\\
IC~1623         &1    &$1.1^{*}$& 0.026   &  2.6        & ---  & 1.0    & 3.32  & 43+43   &  ---\\
CGCG436-30      & --- &         &         & $\lesssim$1 & ---  & ---    & 2.32  & 60      &  ---\\
IRAS~01364-1042 &  1  &$0.8^{*}$& 0.018   &  1.8        & ---  & ---    & 3.44  & 89      &  ---\\
IIIZw35         &  1  &0.86     & 0.031   &  1.67       & ---  & ---    & 2.79  & 72      &  ---\\
UGC~2369        &  1  &0.70     & 0.01    &  0.50       & ---  & $<$0.2 & 2.43  & 63      &  ---\\
IRAS~03359+1523 &  2  &$1.0^{*}$&$0.025^*$&  6.2        & 0.34 & 0.7    & 2.58  & 39+27   & 186 \\%H=40 alpha=1.0
NGC~1614        &  1  &0.63     & 0.017   &  0.24       & ---  & $<$0.6 & 1.81  & 47      &  ---\\
IRAS~05189-2524 &  1  &$0.8^{*}$& 0.030   &  1.4        & ---  & ---    & 3.76  & 97      &  ---\\
NGC~2623        &  1  &$0.8^{*}$& 0.020   &  1.75       & ---  & 0.31   & 2.46  & 49+15   &  ---\\
IRAS~08572+3915 &  1  &$0.8^{*}$&  1      &  4.0        & ---  & 0.31   & 2.73  & 54+17   &  ---\\%PROBLEM SOURCE.
UGC~4881        &  1  &0.75     & 0.033   &  0.32       & ---  & $<$0.48& 2.79  & 72      &  ---\\
UGC~5101        &  1  &$0.8^{*}$& 0.0056  &  1.1        & ---  & 0.54   & 16.9  & 280+150 &  ---\\
IRAS~10173+0828 &  2  &$0.8^{*}$&$0.1^*$  &  (5)        & 0.78 & 2.4    & 2.42  & 18+44   & (110)\\%H fixed at 10
IRAS~10565+2448 &  1  &1.1      & 0.125   &  0.50       & ---  & $<$0.1 & 4.88  & 130     &  ---\\
Arp~148         &  1  &0.68     & 0.026   &  0.21       & ---  & $<$1.1 & 2.85  & 74      &  ---\\
UGC~6436        &  1  &1.1      & 0.1     &  0.60       & ---  & $<$0.04& 1.21  & 31      &  ---\\
Arp~299         &  2  &$1.0^{*}$&$0.125^*$&  (4.2)      &(0.37)& 1.0    & 6.70  & 87+87   & 140 \\ %H=40 alpha=1.
IRAS~12112+0305 &  2  &$0.8^{*}$&$0.1^*$  &  2.1        & 0.3  & 1.2    & 10.2  & 120+140 & 72  \\
UGC8058         &  2  &$0.8^{*}$&$0.1^*$  &  3.9        & 0.21 & 0.52   & 67.4  & $\rm AGN^{\dagger}$+600& 209 \\
UGC~8387        &  2  &$1.4^{*}$&$0.01^*$ &  (3)        &(0.36)& 0.2    & 3.91  & 84+17   & (15)\\%H fixed at 100
NGC5256         &  1  &0.87     & 0.026   &  0.19       & ---  &        & 3.86  & 100     &  ---\\%Straight line?
UGC~8696        &  1  &$0.8^{*}$& 0.067   &  1.0        & ---  & 1.3    & 12.9  & 150+190 &  ---\\
IRAS~14348-1447 &  1  &$0.8^{*}$& 0.0061  &  1.0        & ---  & $<$0.3 & 13.8  & 360     &  ---\\
IZw~107         &  1  &$1.4^{*}$&(0.0067) & (3)        & ---  & 0.1     & 4.71  & 110+11  &  ---\\%alpha=1.4  
IRAS~15250+3609 &  1  &0.84     & 0.024   &  2.2        & ---  & ---    & 6.01  & 160     &  ---\\
Arp~220         &  2  &$0.8^{*}$&$0.1^*$  &  2.0        & 0.32 & 0.78   & 10.8  & 160+120 & 40 \\
NGC~6286        &  1  &0.90     & 0.019   &  0.23       & ---  & $<$1.2 & 2.43  & 63      &  ---\\
NGC~7469        &  1  &$1.0^{*}$& 0.022   &  2.0        & ---  & 0.52   & 2.65  & 45+23   &  ---\\
IC~5298         & --- &         &         & $\lesssim$1 & ---  & ---    & 1.20  & 31      &  ---\\
Mrk~331         & --- &         &         & $\lesssim$1 & ---  & ---    & 1.39  & 36      &  ---\\
\hline
\end{tabular}
\label{tab:turn}
\end{table*}

\begin{table*}
\caption{Emission measures.}
\centering
\begin{tabular}{c c c c c}
\hline\hline
Source       & $\rm EM_1$ & $n_1$ & $\rm EM_2$ & $n_2$ \\
             & $\rm 10^6\;cm^{-6}\,pc$ & $\rm cm^{-3}$ & $\rm 10^6\;cm^{-6}\,pc$ & $\rm cm^{-3}$ \\
\hline
      NGC34 &$\lesssim$3& ---      & ---  & --- \\
     IC1623 & 23        & 330      & ---  & --- \\
 CGCG436-30 &$\lesssim$3& ---      & ---  & --- \\
I01364-1042 & 11        & 290      & ---  & --- \\
    IIIZw35 & 8.2       & 310      & ---  & --- \\
    UGC2369 & 0.8       & 100      & ---  & --- \\
I03359+1523 & 140       & 1400     & 0.32 & 68  \\
    NGC1614 & 0.24      & ---      & ---  & --- \\
I05189-2524 & 6.2       & 200      & ---  & --- \\
    NGC2623 & 9.9       & 290      & ---  & --- \\
I08572+3915 & 56        & 780      & ---  & --- \\
    UGC4881 & 0.32      & 47       & ---  & --- \\
    UGC5101 & 3.7       & 190      & ---  & --- \\
I10173+0828 & 90        &$\geq$1200& 1.8  & $\geq$166 \\
I10565+2448 & 0.40      & 35       & ---  & --- \\
     Arp148 & 0.20      & 42       & ---  & --- \\
    UGC6436 & 0.66      & ---      & ---  & --- \\
     Arp299 & 62        & 900      & 0.38 & 70  \\
I12112+0305 & 15        & 310      & 0.24 & 40  \\
    UGC8058 & 53        &$\geq$950 & 0.12 & $\geq$44  \\
    UGC8387 & 31        & ---      & 0.36 & --- \\
    NGC5256 & 0.074     & 24       & ---  & --- \\
    UGC8696 & 3.1       & 130      & ---  & --- \\
I14348-1447 & 3.1       & 120      & ---  & --- \\
     IZw107 & 31        & 460      & ---  & --- \\
I15250+3609 & 16        &$\geq$500 & ---  & --- \\
     Arp220 & 13        & 410      & 0.28 & 59  \\
    NGC6286 & 0.14      & ---      & ---  & --- \\
    NGC7469 & 13        & 560      & ---  & --- \\
     IC5298 &$\lesssim$3& ---      & ---  & --- \\
     MRK331 &$\lesssim$3& ---      & ---  & --- \\
\hline
\end{tabular}
\label{tab:EM}
\end{table*}

\section{Results}

\subsection{Low frequency radio spectra}

The radio spectra shown in figure.~\ref{fig:fits} show various forms at low frequencies. These include an 
almost straight power-law from 22~GHz down to 244~MHz (e.g. IC~1623), a single turn-over at low frequencies 
(NGC~1614), evidence of two turn-overs (e.g. IR~03359+1523) and a gradual flattening 
towards low frequencies (e.g. NGC~2623). In addition, IR~08572+3915 shows an almost flat spectrum. 
This may rise at frequencies below 1~GHz but the data point at 610~GHz has a large error and 
is marginally consistent with a spectrum that remains flat.     

The flattening of the radio spectra is due to absorption of the power-law synchrotron emission, 
$S_{\nu}\propto \nu^{-\alpha}$, where $\nu$ is frequency. In all but one case this must be free-free 
absorption by ionized gas in the sources because the  
brightness temperatures are not high enough for synchrotron self-absorption to be important. 
Only UGC~8058\footnote{Although this source is known to contain compact radio emission from an AGN
which shows variability on timescales of years, Ulvestad et al. (1999) show that this variability
is likely confined to frequencies above 5~GHz. In our spectrum the data points at 1.4, 4.8, 8.4, 
15 and 22.5 GHz were, in any case, all made on the same date and a turnover is already seen 
between 1.4 and 4.8~GHz. See section~\ref{sec:UGC8058}.} has an 8.4~GHz surface brightness sufficiently 
high that synchrotron self-absorption may be important (Condon et al. 1991).
 
The shape of the low-frequency spectrum depends, almost entirely, on the quantity and 
distribution of ionized gas with respect to the source of synchrotron emission. The 
free-free optical depth is, $\tau_{\nu} \propto n_{e}^2\,L\,\nu^{-2.1}$, where $n_e$ is the 
ionized particle density, $\nu$ is the frequency, and $L$ is the path length along the line-of-sight. 
$n_{e}^2\,L$ is the emission measure. Radio spectra will show a turn-over where $\tau_{\nu} = 1$. 
If the frequency of this turn-over is $\nu_{t}$ then the optical depth is given by, 
$\tau_{\nu} = (\nu/\nu_t)^{-2.1}$.

The radio spectra of \emph{normal}, star-forming galaxies have been reviewed in detail by 
Condon (1992). The spectra are the result of emission from a medium containing both synchrotron
and free-free emitting gas. Based on observational results, Condon assumed that the thermal 
emission contributes 10 times less than synchrotron emission at 1~GHz. Based on equation 6 
of Condon, the radio surface brightness therefore varies with frequency according to,

\begin{equation}
\frac{S}{\Omega} \propto T_e\,(1-e^{-\tau_{\nu}})\,\bigg(1 + \frac{1}{H}\,\bigg(\frac{\nu}{\rm GHz}\bigg)^{0.1-\alpha}\bigg)\,\nu^2
\label{eqn:1}
\end{equation}

\noindent
where $S$ is flux density, $\Omega$ is the area subtended by the source and $T_e$ is the electron 
temperature. $H$ is the thermal fraction. For synchrotron emission that contributes 10 times 
that of thermal emission at 1~GHz, as assumed by Condon (1992), $H=0.1$. Fig. 4 of Condon shows 
the family of curves that result from varying the turn-over 
frequency, the frequency at which the optical depth $\tau_{\nu}=1$. In fig.~\ref{fig:varyH} we 
illustrate the effect of varying $H$ for a fixed turn-over frequency. 

Because we do not have measurements of the the source size, $\Omega$, at every measured 
frequency equation~\ref{eqn:1} is only applicable under the assumption that the size of each 
source is not a function of frequency. Although cosmic ray diffusion is known to make this 
untrue for some nearby objects, the very short synchrotron lifetimes expected in the present 
sample should ensure very similar source sizes at all frequencies under study here.

For normal galaxies the spectral index is close to $-\alpha$ at intermediate frequencies 
(5~GHz) and flattens towards higher frequencies due to the increasing contribution of
free-free emission. Towards low frequencies the spectra turn-over due to the effects of 
free-free absorption. The frequency at which this turn-over occurs depends on the emission
measure of the ionized gas. As pointed out by Condon, real sources, such as M~82, may be 
expected to show a flattening of the radio spectrum towards low frequencies rather than a 
turn-over due to source inhomogeneity.     

We use equation~\ref{eqn:1} as a starting point for the analysis of our radio spectra of 
luminous IRAS galaxies.

Each turn-over frequency in our spectra, and some have more than one, is the result of 
absorption by ionized gas with a different emission measure. Therefore, by fitting one 
or more components described by equation~\ref{eqn:1} we can determine $n_e^2\,L$ for 
our sample galaxies via the standard expression for the free-free absorption coefficient 
(equation 1 in Condon 1992). This step requires an assumption for the electron temperature,
which we take to be $10^4\;\rm K$. In fitting equation~\ref{eqn:1} to the data we consider 
the following cases.

\begin{enumerate}
\item The spectrum is straight to the lowest measured frequency (e.g. IC~5298)- no turnover 
      can be estimated.
\item The spectrum shows a single turn-over. Either only a single absorbed component is 
      visible (e.g. NGC~1614) or there is evidence of another component whose flux continues to 
      rise to the lowest measured frequency (e.g. UGC~5101). In this case we fit equation~\ref{eqn:1} allowing 
      $\alpha$, $H$ and the turn-over frequency to vary. If data points are insufficient we fix 
      $\alpha=0.8$. We allow for the presence of
      an unabsorbed component by fitting an additional power-law with the same power-law index.  
      For those objects where only one component is seen we provide a limit to the possible 
       contribution of an unabsorbed component by finding the maximum contribution of a straight 
       power-law that is consistent with the lowest frequency data point.
      We refer to this as model 1.    
\item The spectrum shows 2 clear turn-overs (e.g. IRAS~03359+1523). We fit 2 components obeying 
      equation~\ref{eqn:1} but
   fix $H=0.1$ and $\alpha=0.8$. For each component, we fit the 
   turn-over frequency and the flux ratio of the components. If equation~\ref{eqn:1} is 
   represented by $f_{i}$ then fit $f = f_1 + x\,f_2$. In this case, if component $f_2$ has 
   the lower break frequency the constant $x$ represents the relative area covered by 
   $f_2$, ie. $\Omega_2/\Omega_1$ because we measure integrated 
   fluxes rather than surface brightnesses. We refer to this as model 2.
\end{enumerate}

Equation~\ref{eqn:1} was fit to the data via a non-linear least squares 
procedure that minimized Chi-squared between the data and model.  
We limited the number of parameters that were allowed to vary so as to
remain fewer than the number of data points.  In table~\ref{tab:turn}
we list the model used and associated fit parameters.  In some cases,
where parameter values were fixed, reasonable fits could not be
obtained with a power-law index of $\alpha=0.8$ and/or $H=0.1$. In
these, alternative fixed values were used, as indicated in
table~\ref{tab:turn}.  

If the best fit solution passed though less than 3/4 of the data 
points, the fit was deemed to be poor. In these cases, shown as dashed 
lines in fig.~\ref{fig:fits}, the derived quantities are uncertain and
appear in parentheses in table~\ref{tab:turn}. In these cases, a good 
fit would require either a more complex model, or more parameters to be 
varied within the adopted model, beyond those permitted by the data.

The model fits are shown in fig.~\ref{fig:fits}.  We note that the
values of $H$, and to a lesser extent $\alpha$, are often rather
poorly constrained by the data.  $H$ tends to be poorly determined
because it is mostly constrained by high frequency data points beyond
8.4~GHz. Here the fractional  thermal emission should be significant
(in normal galaxies $\simeq$30\% of the emission at 8.4~GHz is
thermal). Equation~\ref{eqn:1}  is a spectrum that flattens towards
higher frequencies. However, as already noted in Clemens et al. (2008)
the radio spectra  of LIRGs and ULIRGs often do not show the
flattening expected due to free-free emission and frequently show a
\emph{steepening}. Indeed, Fig.~\ref{fig:fits} has several examples
where the model over-predicts the  flux at $>10\;\rm GHz$. The fitted
thermal fractions, which are generally much lower than 0.1, are a
consequence of the fact  that only a spectral flattening is permitted
by equation~\ref{eqn:1}. This shows that there is an effect that is
not taken into  account in the simple model that we apply here and is
discussed in section~\ref{sec:wherethermal}.

The spectral index, $\alpha$, on the other hand, is poorly constrained by the limited number of data points when the turn-over frequency occurs at frequencies so high that no part of the spectrum is approximately straight.

However, we are concerned mainly with identifying the frequency of the turn-over seen towards longer wavelengths, and the relative strength of the two emission components where a second is present. These values are only very weakly affected by the value of $H$ and weakly by the value of $\alpha$.

\subsection{General form of the radio spectra}

In almost all cases, our low frequency radio data have revealed the presence of a turn-over in the radio spectrum. Although the turn-overs are often at frequencies above 610~MHz, such turn-overs were previously detected only at a single frequency. For an object with only 3 or 4 measured fluxes there was little justification for fitting anything other than a straight line to the data. 

The addition of low frequency radio data to the spectra of this sample of LIRGs and ULIRGs shows the spectra are frequently more complex than previously thought. Galaxies that showed evidence of turn-overs above 1.4~GHz, and would have been expected to show fluxes that fell with decreasing frequency, often show spectra that actually rise again (e.g. NGC~2623). Although Condon (1991) pointed out that free-free absorption in real sources is likely to result in rather broad turn-overs or just a flattening of the spectrum due to source inhomogeneity, we often see rather more complex spectra, indicative of at least 2 emission components. This was unexpected.  

Given our limited sensitivity at 244 and 610~MHz  as well as the fact that there are rarely data below 244~MHz it is likely that all sources in our sample have more than one emission component; we either don't detect the emission or have not observed at sufficiently low frequencies.  

The presence of radio emission components with different characteristic turn-over frequencies tells us that there are populations of relativistic electrons in environments with diverse emission measures. There is more than one physical scenario consistent with this result, and we will discuss this issue in section~\ref{sec:relation}, but the most simple explanation may be that there are starbursts occurring in two or more nuclei (as most of these galaxies are probably merger remnants this would seem plausible). If either the mean ionized gas density or the path length through the ionized gas to the supernovae were different in the two nuclei, different turn-over frequencies would result. Reference to the 8.4~GHz, {0}\farcs{25} resolution maps of Condon et al. (1991) shows that although some sources with 2 emission components do indeed show 2 spatially resolved nuclei (e.g. Arp~220) there are others that do not (e.g. NGC~2623) and also sources that show a single component radio spectrum but have 2 radio-bright nuclei (e.g. IRAS~14348-1447). Of course, higher resolution data may reveal additional double nuclei, and additional radio data at lower frequencies may reveal second emission components where only one is currently detected. Therefore, we cannot, at present, identify separate emission components with separate starburst nuclei.

For all objects with sufficient data we provide the 8.4~GHz flux ratio of the 2 model 
components in table~\ref{tab:turn}, or an upper limit where only a single component is detected.

\subsection{Star formation rates}

In order to estimate the star formation rates associated with each emission component we make use of the calibration of Panuzzo et al. (2003) (based on Bressan et al. 2002). This relates the 8.4~GHz luminosity to the star formation rate for stars with a Salpeter IMF and masses in the range 0.15-120$\;\rm M_{\odot}$ via $SFR = 258.4\times 10^{-23} L_{8.4} \;\rm M_{\odot}yr^{-1}(WHz^{-1})^{-1}$. In table~\ref{tab:turn} we give the observed 8.4~GHz luminosities and implied star formation rates. For those objects with 2 emission components we show the star formation rate for each. For galaxies in which 2 emission components are clearly detected, both imply high star formation rates. Although a high star formation rate is a requirement for detection at frequencies below 1.4~GHz if the radio emission is derived from supernova explosions, the significant fraction of sources in which 2 such components are detected (13/23 with data below 1.4~GHz) means that this is not uncommon among such galaxies. The radio emission at low frequencies is not therefore from a small number of supernovae occurring in more quiescent regions beyond the main starburst, but is rather from a population of supernovae of similar size taking place in a lower density medium (with a lower emission measure).

In those objects where only a single component is detected, the upper limits on the fractional contribution of a component with a lower frequency turn-over only constrains the star formation rate to be $<$10\% that of the main component in 3 cases. The radio emission from infrared luminous galaxies is therefore consistent with high star formation rates in regions with 2 distinct emission measures (but see section~\ref{sec:implications}).  

\subsection{UGC~8058}
\label{sec:UGC8058}

UGC~8058 (Mrk~231) is the only source in our sample with a radio loud
AGN. Vega et al. (2008) found that the contribution of the AGN to the
bolometric luminosity was 24\%. The AGN is also variable, although the
variability is seen only above  5~GHz (Ulvestad et al.,
1999). Fig.~\ref{fig:fits} shows that our model succeeds in fitting
the radio data for this  source perfectly well as two absorbed
components. The discontinuity is probably due to radio emission from
the AGN  that is seen only above 4.8~GHz due to strong absorption. In
this case the absorption may be synchrotron  self-absorption or
free-free absorption. The spectrum implies that, at the epoch of
observation of the data above 1.4~GHz  (1995 November 17), the AGN
contributed 66\% to the radio emission at 8.4~GHz (see
table~\ref{tab:turn}). The radio  emission at lower frequencies is
therefore derived almost entirely from star formation, and we can use
this fact to  estimate the star formation rate. In
table~\ref{tab:turn} we give the star formation rate for this
component alone.  The models of Vega et al. (2008) assumed that AGN
did not emit in the radio. Interestingly, the model for this source
fitted well the low frequency radio data, but under-predicted the
radio emission above 5~GHz.

\subsection{Emission measures}
\label{sec:EM}

Table~\ref{tab:EM} shows the emission measures derived from the fitted
turn-over frequencies listed in table~\ref{tab:turn}.  Also shown are
the mean ionized gas densities implied by calculating the physical source 
sizes via the angular sizes measured at 8.4~GHz and distances given 
by Condon et al. (1991). Our  observations rarely resolve the
sources so we do not have a direct measure of the source sizes at 244
and 610~MHz. These could , in principle, be larger than at 8.4~GHz
due to cosmic ray diffusion, but the synchrotron lifetime is anyway so
short that the effect on source size should be negligible. We cannot
exclude the possibility, however, that the low-frequency  radio
emission comes from a more extended distribution of
supernovae. However, such an extended component would also emit at
8.4~GHz and should have been detected by Condon et al. as extended
halo emission. It is therefore probably reasonable to use  the source
sizes as measured by Condon et al. (1991).

Free-free absorption towards individual supernova remnants has been used to derive emission measures in M~82 (Wills et al. 1997). A large fraction have turn-overs that are evident at 1.4~GHz, implying emission measures $\geq 10^6\;\rm cm^{-6}\,pc$. Other studies have identified regions with emission measures as high as $4\times 10^6\;\rm cm^{-6}\,pc$ (Seaquist, Bell \& Bignell, 1985) and Clemens \& Alexander (2004) derived a value of $10^7$ for the centre of the luminous IRAS galaxy, UGC~8387\footnote{The 2 values we derive here for UGC~8387 straddle this value.}, based on high resolution radio data. Emission measures as high as $1.8\times 10^8\;\rm cm^{-6}\,pc$ have been found for compact sources in the ``Antennae'' galaxies by Neff \& Ulvestad (2000). For those objects which were observed at 244 and 610~MHz we find emission measures in the range $0.07 - 140\times\rm 10^6\;cm^{-6}\,pc$.   

\begin{figure}
\centerline{
\includegraphics[scale=0.45]{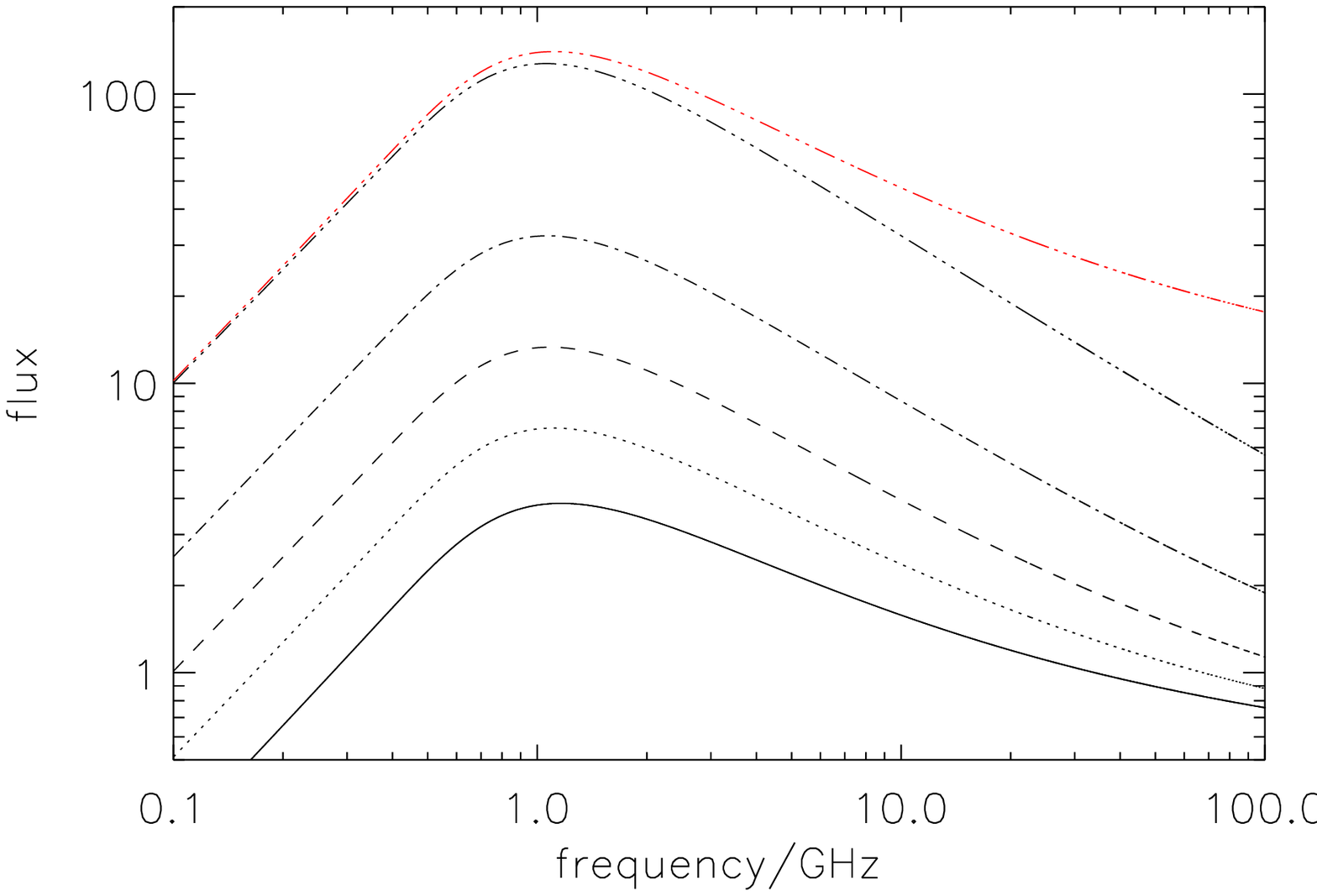}
}  
 \caption{Comparison between varying the thermal fraction and the source size for a given turn-over frequency.
The black lines are for a fixed source size and thermal fractions, $H$, in equation~\ref{eqn:1} of  0.2, 0.1, 
0.05, 0.02 and 0.005, from bottom to top. In this case, increasing flux corresponds to increasing surface brightness. 
The coloured line (top line in plot) has a thermal fraction of $H=0.1$ and has been scaled by $H/0.005$. This corresponds to increasing
the source area by a factor of 20 at fixed thermal fraction. The surface brightness remains constant in this case.}
\label{fig:varyH}
\end{figure}

For those sources in which 2 absorbed components are seen we have fixed the thermal fraction of both so that $H=0.1$. In this case, the spectra imply that the less absorbed component covers a much larger area than that of the more absorbed component, because equation~\ref{eqn:1} implies lower brightness temperatures for lower turn-over frequencies. Because we measure integrated fluxes, a larger source area is needed to fit the data. Of course, this may not reflect the true source structure. The component whose turn-over lies at a lower frequency may instead have a much lower thermal fraction. The similarity between decreasing the thermal fraction by a factor of, say 10, and increasing the source size by a factor of 10 for a given turn-over frequency is shown in fig.~\ref{fig:varyH}. At frequencies near the turn-over, and below, the resulting spectra are rather similar. At higher frequencies there is a significant difference, as the thermal emission contributes a larger fraction of the total. There is little difference, then, for the low frequency end of the model fits between these 2 possibilities, although at higher frequencies we limit our ability to fit the data. However, as mentioned above, the high frequency data are poorly modeled, and so constraining the thermal fraction at $\nu \gtrsim 10\;\rm GHz$ is unreliable in any case. 

To summarize, the relative strength of the two emission components either places a constraint on the relative physical area of each (thermal fraction of both fixed) or places a constraint on the difference in thermal fraction between the two. A combination of these two extremes is also permitted, of course. In table~\ref{tab:turn} we give the value of $x$, which is the relative physical area covered by the less absorbed component, or, almost equivalently, the factor by which this component is poorer in thermal emission. For example, the component with a lower frequency turn-over in Arp~220 is either more spatially extended by a factor of 40 than that of the component with the higher frequency turn-over, or it has a thermal fraction 40 times lower.

\begin{figure}
%\vspace{-5cm}
\hspace{1cm}
\centerline{
\includegraphics[scale=0.5]{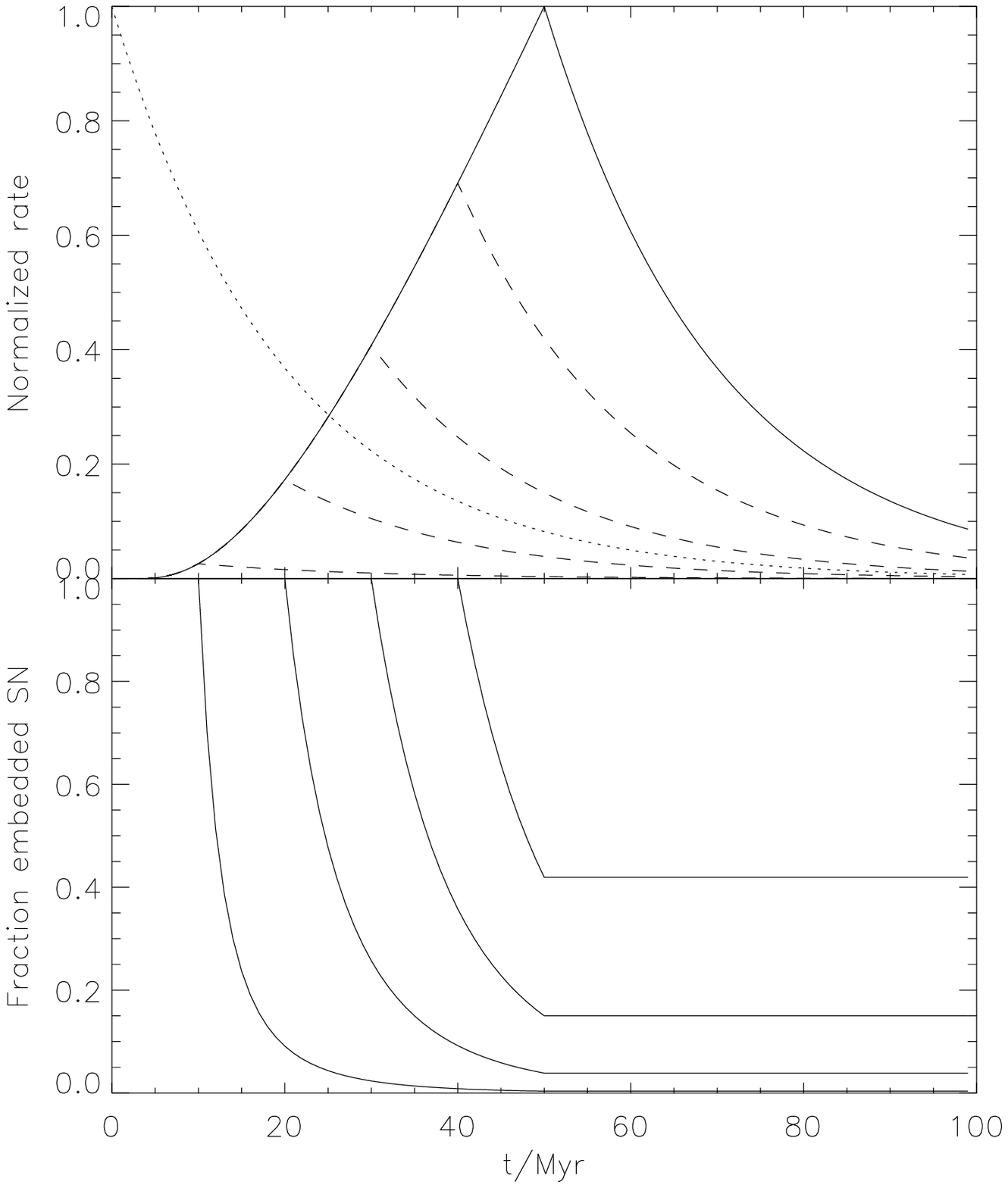}
}  
%\vspace{5cm}
 \caption{{\bf Top:} Supernova rate as a function of time for a star formation rate that decays exponentially and a Salpeter IMF. The star formation rate has an e-folding time of $t_b = 20\;\rm Myr$ and is shown as a dotted line. It is assumed that stars less massive than $8\;\rm M_{\odot}$ do not produce type II supernovae; these have a lifetime $\sim 50\;\rm Myr$. The dashed lines show the supernova rate for stars still embedded in HII regions for various HII region lifetimes. From bottom to top these are, $t_{\rm HII}=(10,20,30,40)\;\rm Myr$. {\bf Bottom:} The fraction of the total supernova rate occurring within HII regions for the above values of $t_{\rm HII}$. We equate this fraction with the observed fractional radio emission of the more and less absorbed components, see table~\ref{tab:turn}.}
\label{fig:SNR}
\end{figure}

\begin{figure}
\centerline{
\includegraphics[scale=0.38]{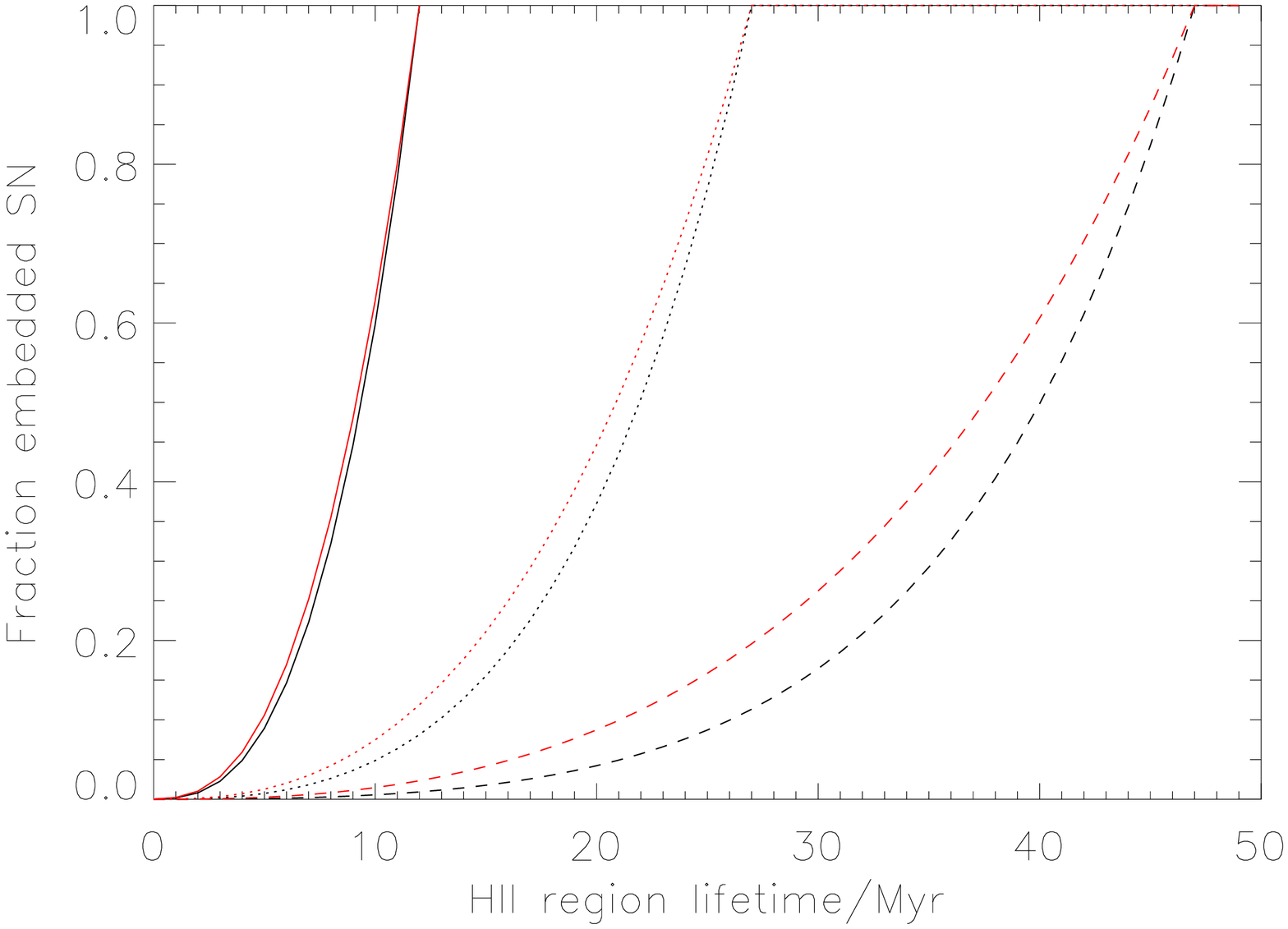}
}  
 \caption{Fraction of supernovae occurring within dense HII regions as a function of the HII region lifetime, $t_{\rm HII}$. The star formation rate decays exponentially, $\psi(t)\propto e^{-t/t_b}$. The solid, dotted and dashed lines are for ages, $t_{\rm age} = (15, 30, 50)\;\rm Myr$. The black curves are for e-folding time, $t_b=15\;\rm Myr$ while the coloured lines are for $t_b=30\;\rm Myr$.}
\label{fig:models2}
\end{figure}

\begin{figure}
\centerline{
\includegraphics[scale=0.38]{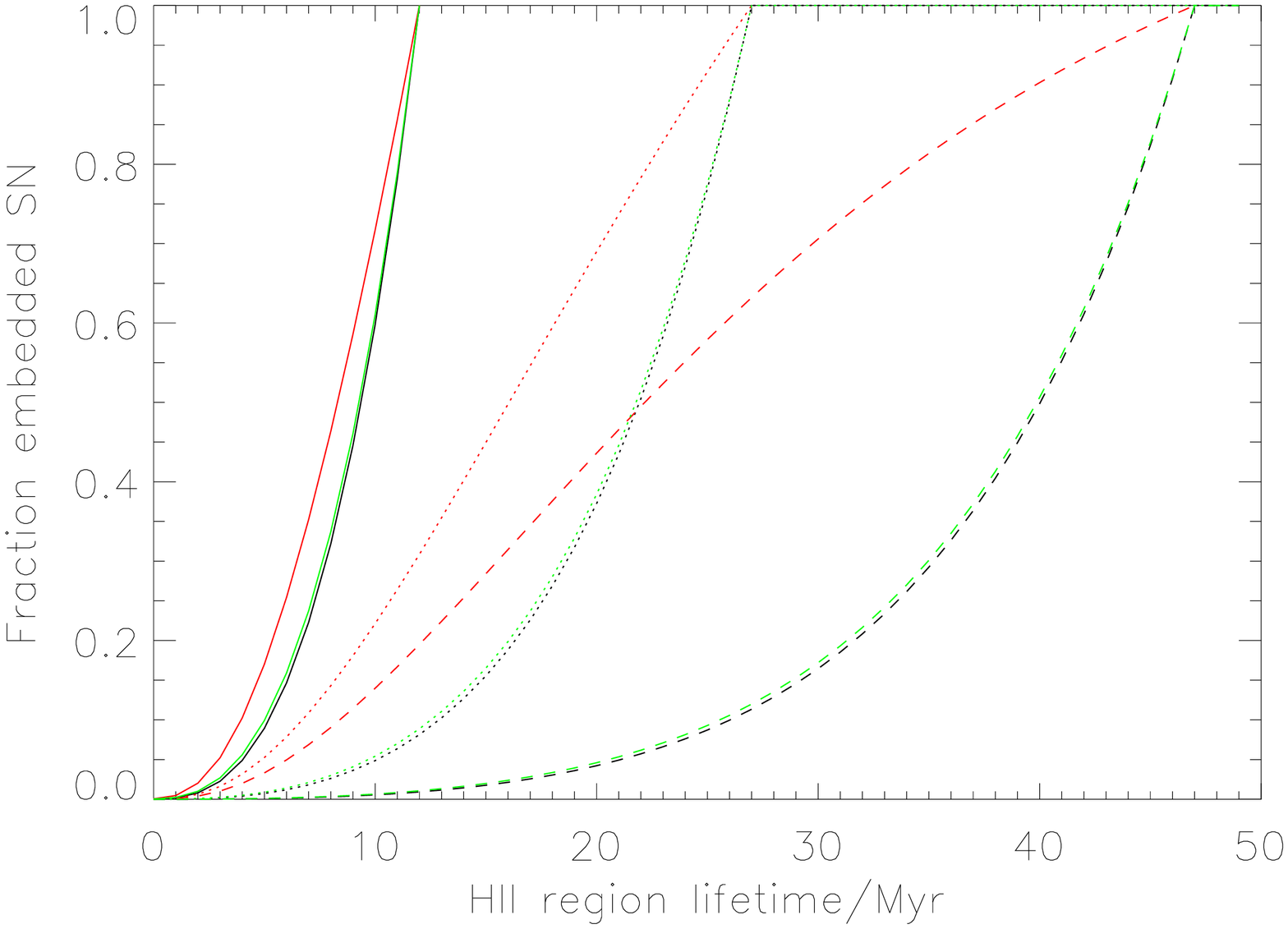}
}  
 \caption{Effect of a rising star formation rate. The black curves are the same as those in fig.~\ref{fig:models2} while the red curves are for a starburst that rises in time as $\psi(t)\propto e^{t/t_b}$. The green lines show the effect of a flatter IMF for high mass stars ($dn/dM \propto M^{-2.15}$, Baldry \& Glazebrook, 2003) for an exponentially decreasing burst.}
\label{fig:models3}
\end{figure}

\section{Implications for starburst evolution}
\label{sec:implications}

As discussed in Condon et al. (1991), with the exception of UGC~8058, there is 
little evidence that any of the galaxies in our sample contain a radio loud AGN. Vega et al. (2008), 
in fact, demonstrate that star formation alone is sufficient to explain the radio emission of every 
source except UGC~8058. In what follows, we assume that the radio emission is derived entirely from
processes linked to star formation.

In this section we focus particularly on the presence of two radio emission components in many sources 
in order to investigate the structure and evolution of the starbursts in these galaxies.

\subsection{Relation between supernovae and ionized gas}
\label{sec:relation}

As we have seen, the form of the radio spectra shown in Fig.~\ref{fig:fits} imply that synchrotron
emission commonly emerges from at least two distinct regions with quite different emission measures.
Such a situation could arise either because supernovae occur in two different environments or because
cosmic rays are transported from one environment to the other after their acceleration in supernova
shocks. Because higher energy electrons lose energy to synchrotron and inverse Compton losses more 
rapidly than low energy electrons, the low energy electrons emit synchrotron photons for longer and 
can therefore be transported further from their source. If all supernovae occur in dense ionized media, 
consistent with regions of higher emission measure, the lower energy electrons may be transported 
to regions of lower emission measure while the more energetic particles end their synchrotron emitting 
lives still within the dense gas.

If the transport of relativistic electrons within HII regions is well-described by a diffusion coefficient, 
$D$, then they can be transported a distance $d\sim (D\,\tau)^{1/2}$ before ceasing to emit synchrotron
radiation (Condon 1992). Although we do not know the value of $D$ in our sources, it has been estimated for 
the starburst galaxy NGC~2146 by Lisenfeld et al. (1996) to be in the range $1-4\times 10^{28}\;\rm cm^{2}\,s^{-1}$.
If relativistic electrons emitting at 244~GHz have a synchrotron lifetime, $\tau_s \simeq 5\times 10^4\;\rm yr$ 
in the strong magnetic fields characterizing these sources, then
they could be transported a distance $d \sim 80\;\rm pc$. As giant HII regions have sizes of order 10~pc and 
relativistic electrons may be transported by methods more efficient than diffusion (outflows or winds) the escape 
of relativistic electrons from HII regions seems feasible.

However, our radio spectra are characterized by quite distinct components, as opposed to a gradual changes in 
spectral index. Because the synchrotron lifetime is only a weak function of the emission frequency, 
$\tau_s \propto \nu^{-1/2}$, the difference in lifetime of electrons emitting at 8.4~GHz and at 244~MHz is 
only a factor of 6 for a given magnetic field strength. Given the inevitable inhomogeneity of these sources on 
parsec scales it seems extremely unlikely that distinct spectral breaks could be caused by the preferential escape
of lower energy cosmic rays.      

It is far more likely, in fact expected, that supernovae occur in different environments. If the supernovae
responsible for the synchrotron emission are those with progenitors with masses $\geq 8\;\rm M_{\odot}$ (Portinari, 
Chiosi \& Bressan 1998) then synchrotron emission is generated for $\sim 50\;\rm Myr$ (for solar metallicity)
after a burst of star formation, ie. the lifetime of an $8\;\rm M_{\odot}$ star. Massive stars are born in dense 
molecular clouds, which they then ionize once luminous. 
Once these stars have formed, the parent cloud immediately stars to be disrupted, first by stellar winds and then 
by the first supernova explosions, which will occur after $\sim 3$ million years for a $100\;\rm M_{\odot}$ star. 

The synchrotron emission component that shows the higher frequency turn-over, implying a higher emission measure, 
should therefore originate from cosmic rays accelerated by the most massive stars, that end their lives while 
still embedded in their parent HII regions. The less absorbed synchrotron component has its origin in supernova
explosions of less massive stars (down to $8\;\rm M_{\odot}$) that out-live the parent HII region.

In those sources with 2 emission components the lifetime of HII regions is therefore bounded, on the low side, by
the lifetime of the most massive stars, $\sim 3\;\rm Myr$, and on the high side by the lifetime of an $8\;\rm M_{\odot}$
star. Only if HII region lifetimes lie in this range will emission components with very different emission measures be 
seen. For sources in which the burst of star formation began more recently than 50~Myr ago the constraint on HII region
lifetime becomes,  $3\;\rm Myr \lesssim t_{\rm HII} \lesssim t_{age}$. The observed fraction of emission in each 
radio component constrains the lifetime within this range. Below, we attempt to estimate the mean HII region lifetime
from the observed spectra.       

\subsection{HII region lifetimes}

The HII region lifetime derived from the relative strength of the two radio emission components depends on various 
starburst parameters that are not known.
These are: the IMF, the temporal evolution of the star formation rate and the synchrotron emission from a 
supernova as a function of progenitor mass. Bearing these limitations in mind, we can nonetheless make use of the
modeling carried out in Vega et al. (2008). In this work, we modeled the spectral energy distribution from the
 near infrared to the radio of the present sample of LIRGs and ULIRGs with starburst models. For each object, both 
the starburst age, $t_{\rm age}$, and the e-folding time of an assumed exponentially decaying burst, $t_b$, were derived. 

Because the starburst timescales in these objects are thought to be very short, ie, of the the same order as the 
lifetime of the stars that undergo supernovae, we cannot assume that the supernova rate bears a one-to-one relation
to the star formation rate. After the start of a burst, that then decreases in intensity with time, the supernova rate 
will initially rise. We estimate the supernova rate as follows. 
If the mass of a star and its lifetime are related via $M(t)$, then for an IMF, $\psi(M)$, the number of stars as 
a function of their lifetime is $\psi(M(t))$. If a starburst occurred at time $t=0$, then at later time, $t_{age}$,
the number of stars about to end their lives is $\psi(M(t_{age}-t))$. The total supernova rate at time $t_{age}$ is 
then given by,

\begin{equation}
SNR = \int_{t1}^{t_{\rm age}-t_{\rm min}} \phi(t)\,\psi(M(t_{\rm age}-t))\,dt
\label{eqn:2}
\end{equation}

\noindent
If $t_{\rm min}$ and $t_{\rm max}$ are respectively the lifetimes of
the most massive and least massive stars that  end their lives as type
II supernovae ($t_{\rm min}\sim 3\;\rm Myr$ for a $100\;\rm M_{\odot}$
star and  $t_{\rm max}\sim 50\;\rm Myr$ for an $8\;\rm M_{\odot}$
star), then if $t_{\rm age} \leq t_{\rm max}$ then $t1=0$, but if the
burst is older than the lifetime of the longest lived supernova
precursors, so that $t_{\rm age} > t_{\rm max}$, then  $t1 = (t_{\rm age}-t_{\rm max})$. 
$\phi(t)$ is the star formation rate. 

Some fraction of the supernovae, occurring at the time dependent rate
given by equation~\ref{eqn:2}, will explode while still embedded in
their parent HII regions. These will be those stars with lifetimes
shorter than the HII  region lifetime, $t_{\rm HII}$. In
equation~\ref{eqn:2} the lower limit on the integral becomes
$(t_{age}-t_{\rm HII})$  in this case.  

Because the synchrotron emitting lifetime of relativistic electrons,
$\sim 10^4-10^5$ years in our frequency range, is much shorter than
any timescale that characterizes the starbursts, we can interpret this
supernova rate directly as  the radio emission, if we assume that the
radio emission per supernova is independent of the progenitor mass. 

We estimate $M(t)$ from an analytical fit to the solar metallicity
stellar evolution models of Bressan et al. (1993).

\begin{equation}
\log M = 1.389 -0.351\log(t) + 0.223\log(t)^{-2}
\end{equation}

We assume a Salpeter IMF, $\psi(M)\propto M^{-2.35}$, and
exponentially decreasing star formation rates,  $\phi(t)\propto e^{-t/t_{b}}$, 
where $t_{b}$ is the burst e-folding time.
Figure~\ref{fig:SNR} illustrates how the total supernova rate varies
as a function of time under these assumptions.  The sharp peak is
caused by the assumption that stars less massive than $8\;\rm M_{\odot}$ 
do  not produce type II supernovae; these have a lifetime
taken to be $50\;\rm Myr$.\footnote{If stars with masses as  low as,
 say, $5\;\rm M_{\odot}$ underwent type II supernovae the supernova
  rate would be unchanged up until  $50\;\rm Myr$ after the start of a
  burst. Thereafter, the supernova rate would continue to increase
  until  $\sim 100\;\rm Myr$, the lifetime of a $5\;\rm M_{\odot}$
  star. The supernova rate peak would occur at $100\;\rm Myr$  rather
  than at $50\;\rm Myr$. In both cases the fraction of embedded
  supernovae reaches a steady state after the  peak. For a given HII
  region lifetime, this steady value is a smaller fraction in the
  $5\;\rm M_{\odot}$ case (lower  mass stars are more numerous and
  live longer, so the supernovae rate would be dominated by stars that
  have out-lived  their HII regions). The effect of this, is that for
  sources with $t_{\rm age} < 50\;\rm Myr$, there is no change in  the
  derived HII region lifetimes, whereas for those with $t_{\rm age} >
  50\;\rm Myr$ the implied HII region lifetimes  would increase.} The
dashed lines show the supernova  rate for stars still embedded in HII
regions for various values of $t_{\rm HII}$. We assume that all massive 
stars are born within HII regions and that after a time, 
$t_{\rm HII}$, these stars are no longer within their parent HII region. Stars with lifetimes
shorter than $t_{\rm HII}$ will explode as supernovae while still within the HII region while
those with lifetimes $> t_{\rm HII}$ will produce supernovae external to HII regions.  

By using the starburst
e-folding  time and starburst ages, as given by Vega et al. (2008), we
can plot, for each source, the time variation of the total  supernova
rate and the fraction of supernovae occurring while still embedded in
HII regions. The fraction of supernovae  occurring in HII regions as a
function of the HII region lifetime, $t_{\rm HII}$, for a given
$t_{\rm age}$ and $t_{b}$ is  shown in  Fig.~\ref{fig:models2}. 

If we assume that the observed radio spectra reflect the properties of the mean HII region then
these
fractions can be compared with the observed radio flux ratio as given
by the values of  $f_2/f_1$ in table~\ref{tab:turn}. $f_1$ then refers to radio emission from 
supernovae occurring within HII regions, while $f_2$ refers to that produced outside HII regions. 
For each source, we therefore find the value of $t_{\rm HII}$, and these are given  in
table~\ref{tab:models}.

\section{Discussion}
\label{sec:discussion}

Individual HII regions cannot have lifetimes as long as most of those given in table~\ref{tab:models}. They can only 
remain ionized for a time equal to the lifetime of the least massive stars with sufficient ionizing flux. This is only 
a few Myr. Therefore, at least for an exponentially decreasing burst, the HII regions that cause the 
free-free absorption cannot be those that host most of the supernovae explosions that generate the radio emission. 
Younger star forming regions, still rich in ionized gas, may lie along the line of sight to a region of 
supernovae however, and cause the free-free absorption. Such a situation would naturally result if star formation were 
initiated in the nuclei of the mergers and then propagated outwards, perhaps triggered by supernova shocks. In addition, 
this would create a shell-like (foreground screen) geometry for the ionized gas and free-free absorption would be much more 
efficient. Therefore, what we actually constrain is not the lifetime of an individual HII region in the usual sense, but
rather, the time for which ionized gas, sufficient to cause the observed absorption, persists along the line-of-sight. This 
ionized gas need not be in a single `HII region'. In what follows, we refer to `regions of ionized gas' to avoid the 
connotation of a single cloud of gas. 

But how does the situation change if the starburst is in an increasing phase? For a rising star formation rate the required
ionized gas lifetimes will decrease because the current star formation rate, resulting in embedded supernovae, is higher than that 
some Myr in the past, resulting in less embedded supernovae. Figure~\ref{fig:models3} compares the fraction of embedded 
supernovae as a function of $t_{\rm HII}$ for an exponentially \emph{rising} star formation rate, with one that falls 
exponentially, as in fig.~\ref{fig:models2}. It is clear from this, that for a given fraction of embedded supernovae 
(absorbed radio flux), the required values of $t_{\rm HII}$ are indeed much shorter. The rising star formation rate mitigates
the accumulation of supernovae that occur outside regions of ionized gas, that otherwise cause the majority of supernovae to occur
outside such regions. The effects of a rising star formation rate are important if $t_{\rm b}\ll 50\;\rm Myr$. 

Much shorter values of  $t_{\rm HII}$ would also result if the source ages were simply much lower than assumed. However, 
given that the supernova rate does not peak until $t_{\rm age} \gtrsim 50\;\rm Myr$, for almost any star formation rate and 
reasonable IMF, we cannot have a sample for which all the sources are younger than 50~Myr. We would have somehow 
selected out the older sources, that nonetheless would be strong radio emitters. Given that the original sample is 
radio selected (Condon et al. 1991) the derived ages cannot be very much younger than those given in Vega et al. (2008). 

Another possibility is that the synchrotron emission is produced over shorter time scales. This could be achieved if the IMF
were significantly flatter than the Salpeter value for high mass stars, as proposed by Baldry \& Glazebrook 
(2003). However, the magnitude of the effect is very small, as shown in fig.~\ref{fig:models3}. Shorter stellar lifetimes also 
result if source metallicities are higher than solar. For a metallicity 2.5 times the solar value ($Z=0.05$) an  
$8\;\rm M_{\odot}$ star lives for 32~Myr, rather than 50~Myr, Fagotto et al.(1994). 

Until now we have assumed that the free-free absorption is caused by gas that has been ionized by the uv photons of young, 
massive stars. If this were not the case, then the time for which a region can remain ionized is no longer limited by the 
lifetime of massive stars. An AGN can, in principle, ionize large volumes of inter-stellar medium if it emits strongly at 
uv wavelengths and shorter. Vega et al. (2008) estimated the AGN contribution to the bolometric luminosity in the present
sample of infrared luminous sources and found that only 9/30 sources have an AGN contribution that exceeds 10\%. Only 
IRAS~08572+3915 was found to be AGN dominated, with a 52\% contribution from an AGN. Although AGN, where present, may 
contribute to the ionizing flux of a source, it seems unlikely that this would dominate over the contribution from star 
formation, which certainly produces ionizing radiation, and almost always dominates the AGN in terms of bolometric 
luminosity. An AGN that contributed additional ionizing flux, over and above that produced by star formation, would also
create an excess of free-free emission, visible as a flattening of the spectra towards higher frequencies. As mentioned above,
and discussed in the next section, we actually find a deficit of free-free emission with respect to that expected for star 
formation alone.    

Finally, we note that the values that we derive for $t_{\rm HII}$ are similar to the values of the `escape time' found in 
Vega et al. (2008). This escape time parameterizes the time that elapses before stars escape the confines of the molecular 
clouds in which they were formed, and is described in Silva et al. (1998). The escape time is largely constrained by the 
spectral shape in the mid and far-infrared. Both these escape times and the values of $t_{\rm HII}$ that we derive above, 
which are an order of magnitude larger than the values found in normal star forming galaxies, suggest that the star 
formation in these objects is embedded in very large regions of dense gas. The effects of age-dependent extinction are 
thus also seen in the radio regime, as in other parts of the spectrum.

\subsection{Where is the thermal emission?}
\label{sec:wherethermal}

The thermal fractions shown in table~\ref{tab:turn} are almost always lower than the 0.1 assumed by Condon (1992) 
for ``normal'' galaxies. Despite this, there are still many instances where the data lie below the fitted curve. Among these,
there are those that show a \emph{steepening} of the spectra towards higher frequencies (eg. Arp~299), something that even 
a complete absence of thermal emission cannot achieve. The simple model we apply to these sources therefore fails at 
high frequencies. There is some additional effect influencing the radio emission in these luminous infrared galaxies
that is not required to fit the radio spectra of normal galaxies.

The absorption of ionizing photons by dust can reduce the quantity of ionized gas in ULIRGs (Valdes et al. 2005) thus reducing 
the thermal emission. However, although this effect may be present it cannot explain the spectra of those sources for which the
spectra actually \emph{steepen} above $\sim 10\;\rm GHz$. 

As discussed in Clemens et al. (2008), synchrotron aging is not expected to cause a steepening of the radio spectra
towards high frequencies in these sources because the synchrotron lifetime is very short. One actually expects spectra to be 
globally steeper than those of normal galaxies. The fact that this is not the case has been discussed by Thompson et al. (2006).
These authors consider the effects of ionization losses on the radio spectra of starburst galaxies and find that the radio 
spectra gradually flatten towards low frequencies and that the spectral indices in the GHz range are similar to those of 
normal galaxies. If ionization losses are indeed important in these kind of galaxies it seems likely that the effects of
ionization losses could explain why we observe unexpectedly steep spectra at high frequencies in fig.~\ref{fig:fits}.

\begin{table}
\caption{Derived HII region survival times for a Salpeter IMF. $T$ is the lesser of the source age as derived in Vega et al. (2008) and 50~Myr; the lifetime of an $8\;\rm M_{\odot}$ star. $t_{\rm HII}^-$ are the survival times for exponentially decreasing star formation rates, $t_{\rm HII}^+$ for exponentially increasing star formation. In both cases the e-folding time for the star formation in each source is as derived in Vega et al. (2008).}
\centering
\begin{tabular}{c c c c c c}
\hline\hline
Source       & $t_{age}$ & $t_{esc}$ & $T$ & $t_{\rm HII}^-$ & $t_{\rm HII}^+$ \\
             & Myr      & Myr      & Myr & Myr & Myr \\
\hline
      NGC34 & 75 & 65 & 50  & --- & ---  \\%2.52
     IC1623 & 43 & 24 & 43  & 33  & 25   \\%2.14
 CGCG436-30 & 32 & 24 & 32  & --- & ---  \\
I01364-1042 & 19 & 14 & 19  & --- & ---  \\%2.67   -
    IIIZw35 & 35 & 35 & 35  & --- & ---  \\%2.58   -
    UGC2369 & 69 & 36 & 50  &$>$45& $>$40\\%2.39   +
I03359+1523 &--- &    & --  & --- & ---  \\%2.58   - 
    NGC1614 & 69 & 50 & 50  &$>$42& $>$30\\%2.58   +
I05189-2524 &5.6 & 6  & 5.6 & --- & ---  \\%2.76   -
    NGC2623 & 53 & 30 & 50  & 43  & 39   \\%2.51   -
I08572+3915 &3.9 & 3  & 3.9 & 1   & 1    \\%3.11   - 
    UGC4881 & 97 & 44 & 50  &$>$41& $>$36\\%2.50   +
    UGC5101 & 28 & 22 & 28  & 23  & 15   \\%2.10
I10173+0828 & 16 & 14 & 16  & 8   & 8    \\%2.92   -
I10565+2448 & 81 & 60 & 50  &$>$46& $>$44\\%2.55   +
     Arp148 & 54 & 27 & 50  &$>$39& $>$25\\%2.54   +
    UGC6436 & 98 & 34 & 50  &$>$47& $>$46\\%2.57   +
     Arp299 & 57 & 43 & 50  & 39  & 27   \\%2.35   -
I12112+0305 &7.6 & 7  & 7.6 & 4   & 4    \\%2.66   -
    UGC8058 & 37 & 26 & 37  & 31  & 27   \\%2.24   -
    UGC8387 & 52 & 33 & 50  & 45  & 39   \\%2.28   -
    NGC5256 & 43 & 15 & 43  & --- & ---  \\%1.99   +
    UGC8696 & 45 & 37 & 45  & 34  & 22   \\%2.31   +   
I14348-1447 & 15 & 14 & 15  &$>$12& $>$11\\%2.38   
     IZw107 & 59 & 40 & 50  & 46  & 44   \\%2.38   -
I15250+3609 & 19 & 16 & 19  & --- & ---  \\%2.81   -
     Arp220 & 22 & 20 & 22  & 16  & 15   \\%2.63   -
    NGC6286 & 44 & 23 & 44  &$>$36& $>$16\\%2.05   +
    NGC7469 & 87 & 45 & 50  & --- & ---  \\%2.29   -    - 2.5847
     IC5298 & 79 & 60 & 50  & --- & ---  \\%            + 2.3867
     MRK331 & 81 & 35 & 50  & --- & ---  \\%2.51
\hline
\end{tabular}
\label{tab:models}
\end{table} 

\section{Conclusions}

We have combined new measurements of the 244 and 610~MHz fluxes, obtained with the GMRT, of a sample of 20 LIRGs and ULIRGs 
with existing radio data. The resulting radio spectra show a range of bends and turn-overs that can be well explained in 
terms of free-free absorption. In several cases two radio components with different emission measures are necessary to fit 
the data. 

For those objects with two emission components, we estimate the fraction of radio emission associated with each. Via a simple 
model for the temporal evolution of the supernova rate, we estimate the fraction of supernovae occurring inside and 
outside of their parent HII regions as a function of HII region lifetime. For simple exponentially decreasing star formation 
rates the lifetimes so derived are too long for them to refer to single HII regions. Altering the form of the high mass end of 
the IMF or the metallicity has relatively little affect on this result. However, if the star formation rate for these sources 
is currently rising, rather than declining, the derived HII region lifetimes are very much reduced. The effect 
of a rising star formation rate is to increase the number of younger (high mass) supernovae occurring within HII regions, 
relative to the number of older (lower mass) supernovae occurring outside HII regions. The star formation rate rise times 
should be much less than the lifetime of the least massive stars that produce type II supernovae. Despite this reduction,
the derived ages remain longer than those plausible for individual HII regions. Such large ages imply that the physical 
situation is one in which younger HII regions obscure older star formation regions along the line of sight. This situation 
would arise naturally if star formation propagated outwards during a burst. Because a rising star formation rate creates 
conditions more conducive to free-free absorption, and the fact that the number of sources with two emission components 
is close to 50\%, it is likely that those in which only a single emission component is present are those with falling 
star formation rates.  

The existence of two emission components in the radio therefore points to an age-dependent extinction analogous to that 
seen at other wavelengths. In the case of radio emission, the most massive stars, that end their lives earlier, 
 produce synchrotron emission that is subjected to preferentially more free-free absorption than those, lower mass, stars 
that live for longer.

The observed over-prediction of fluxes at frequencies above $\sim 10\;\rm GHz$ may be resolved if the intrinsic 
synchrotron spectrum were not straight, but had a steeper spectral index at higher frequencies due to 
ionization losses, as modeled by Thompson et al. (2006).  

The two radio components in UGC~8058 can be identified with the radio loud AGN and star formation; the former contributing 66\% 
of the radio emission. This interpretation is consistent with the model of Vega et al. (2008).

\section*{Acknowledgments}
MC and AS acknowledge the expert assistance of Ishwara Chandra and the GMRT operators during the observations. 
MC and AB acknowledge support from contrat ASI/INAF I/016/07/0.
This research has made use of the NASA/IPAC Extragalactic Database (NED) which is operated by the Jet Propulsion Laboratory, California Institute of Technology, under contract with the National Aeronautics and Space Administration.

%\begin{thebibliography}{99}

\section{References}

%\bibitem[]{}Baldry, I.~K., Glazebrook K., 2003, ApJ, 593, 258
Baldry, I.~K., Glazebrook K., 2003, ApJ, 593, 258\\
Bressan, A., Fagotto, F., Bertelli, G., Chiosi, C., 1993, A\&AS, 100, 647\\
Bressan A., Silva L., Granato G.~L., 2002, A\&A, 392, 377\\ 
Clemens M.~S., Vega O., Bressan A., Granato G.~L., Silva L., Panuzzo P., 2008, A\&A, 477, 95\\
Clemens M.~S., Alexander P., 2004, MNRAS, 350, 66\\
Condon J.~J., 1992, ARA\&A, 30, 575\\
Condon J.~J., Huang Z.-P., Yin Q.~F., Thuan T.~X., 1991, ApJ, 378, 65\\
Fagotto, F., Bressan, A., Bertelli, G., Chiosi, C., 1994, A\&AS, 104, 365\\
Lisenfeld U., Alexander P., Pooley G.~G., Wilding T., 1996, MNRAS, 281, 301\\
Neff S.~G., Ulvestad J.~S., 2000, AJ, 120, 670\\
Panuzzo, P., Bressan, A., Granato, G.~L., Silva, L., Danese, L., 2003, A\&A, 409, 99\\
Portinari L., Chiosi C., Bressan A., 1998, A\&A, 334, 505\\
Seaquist E.~R., Bell M.~B., Bignell R.~C., 1985, ApJ, 294 546\\
Silva L., Granato G.~L., Bressan A., Danese L., 1998, ApJ, 509, 103\\
Thompson T.~A., Quataert E., Waxman E., Murray N., Martin C.~L., 2006, ApJ, 645, 186\\
Vald{\'e}s J.~R., Berta S., Bressan A., Franceschini A., Rigopoulou D., Rodighiero G., 2005, A\&A, 434, 149\\
Vega O., Clemens M.~S., Bressan A., Granato G.~L., Silva L., Panuzzo P., 2008, A\&A, 484, 631\\
Wills K.~A., Pedlar A., Muxlow T.~W.~B., Wilkinson P.~N., 1997, MNRAS, 291, 517\\

%\end{thebibliography}

\bsp

\label{lastpage}

\end{document}